\documentclass[12pt]{article}

\usepackage{caption,subcaption}

\pdfoutput=1

\usepackage{appendix}

\usepackage{amsfonts,amsmath,amssymb,accents,mathrsfs}
\usepackage[retainorgcmds]{IEEEtrantools}
\usepackage{multirow}
\usepackage{multicol}
\usepackage{tikz}
\usepackage{graphicx,color}
\usepackage{booktabs}
\usepackage{placeins}
\usepackage[colorlinks,linkcolor=Blue,citecolor=Blue,urlcolor=Blue,bookmarks,bookmarksnumbered]{hyperref}

\usepackage[nosort]{cite}

\usepackage[scaled=0.85]{helvet}
\usepackage[T1]{fontenc}
\usepackage[utf8]{inputenc}
\usepackage{cancel}

\definecolor{Red}    {rgb}{0.90,0.00,0.12} %  1
\definecolor{Blue}   {rgb}{0.00,0.00,1.00} %  2
\definecolor{Green}  {rgb}{0.10,0.70,0.10} %  3
\definecolor{Turque} {rgb}{0.00,0.65,0.85} %  4
\definecolor{Orange} {rgb}{1.00,0.50,0.15} %  5
\definecolor{Magenta}{rgb}{1.00,0.00,1.00} %  6
\definecolor{Gold}   {rgb}{1.00,0.75,0.25} %  7
\definecolor{Seaweed}{rgb}{0.01,0.24,0.09} %  8
\definecolor{Purple} {rgb}{0.50,0.25,0.55} %  9
\definecolor{Brown}  {rgb}{0.43,0.26,0.32} % 10
\definecolor{grey1}  {rgb}{0.20,0.20,0.20} % 11
\definecolor{grey2}  {rgb}{0.40,0.40,0.40} % 12
\definecolor{grey3}  {rgb}{0.60,0.60,0.60} % 13
\definecolor{grey4}  {rgb}{0.80,0.80,0.80} % 14
\definecolor{grey5}  {rgb}{0.90,0.90,0.90} % 15

\def\a{{\alpha}}

\def\ad{{\dot{\alpha}}}

\def\N{{\mathcal{N}}}

\def\J{{\mathcal{J}}}

\def\Ysf{{\textsf{Y}}}

\def\D{{\rm D}}
\def\Dd{{\bar{\rm D}}}
\def\pa{\partial}

\def\ff#1-#2{\frac{#1}{#2}}
\def\tff#1-#2{\tfrac{#1}{#2}}

\def\be{\begin{equation}}
\def\ee{\end{equation}}
\def\bea{\begin{eqnarray}}
\def\eea{\end{eqnarray}}
\def\n{\IEEEyesnumber}

\makeatletter
\def\section{\@startsection{section}{1}{\z@}
              {3ex plus-1ex minus-.2ex}{1pt plus1pt}
              {\large\sf\bfseries\boldmath}}
\def\subsection{\@startsection{subsection}{2}{\z@}
              {1.5ex plus-1ex minus-.2ex}{0.01pt plus1pt}{\sf\slshape}}
\def\subsubsection{\@startsection{subsubsection}{3}{\z@}
              {1.5ex plus-1ex minus-.2ex}{0.01pt plus0.2pt}{\sf\boldmath}}
\def\paragraph{\@startsection{paragraph}{4}{\z@}
              {.75ex \@plus.5ex \@minus.2ex}{-2mm}{\sf\bfseries\boldmath}}
\makeatother

   \parskip\medskipamount     \lineskip=0pt
\thicklines
\begin{document}
\thispagestyle{empty}
\noindent{\small
%\hfill{\# HET {~} \\ % un-comment-out and specify when done}
%$~~~~~~~~~~~~~~~~~~~~~~~~~~~~~~~~~~~~~~~~~~~~~~~~~~~~~~~~~~~~$
%$~~~~~~~~~~~~~~~~~~~~\,~~~~~~~~~~~~~~~~~~~~~~~~~\,~~~~~~~~~~~~~~~~$
% {~}
%}
\vspace*{6mm}
\begin{center}
{\large \bf 
Higher Spin Supersymmetry at the Cosmological Collider:\\
Sculpting SUSY Rilles in the CMB
\vspace{3ex}
} \\   [9mm] {\large { 
Stephon Alexander\footnote{stephon\_alexander@brown.edu}$^{a,}$,
S.\ James Gates Jr.\footnote{sylvester\_gates@brown.edu}$^{a}$,
Leah Jenks\footnote{leah\_jenks@brown.edu}$^{b}$,
K.\ Koutrolikos\footnote{konstantinos\_koutrolikos@brown.edu}$^{a}$
and Evan McDonough\footnote{evan\_mcdonough@brown.edu}$^{a}$ }
}
\\*[8mm]
\emph{
\centering
$^{a}$Brown Theoretical Physics Center and Department of Physics, Brown University, Providence, RI 02912, USA
\\[6pt]
$^{b}$Department of Physics, Brown University,
\\[1pt]
Box 1843, 182 Hope Street, Barus \& Holley 545,
Providence, RI 02912, USA 
}
 $$~~$$
  $$~~$$
 \\*[-8mm]
{ ABSTRACT}\\[4mm]
\parbox{142mm}{\parindent=2pc\indent\baselineskip=14pt plus1pt
We study the imprint of higher spin supermultiplets
on cosmological correlators, namely the non-Gaussianity of the cosmic microwave 
background. Supersymmetry is used as a guide to introduce the contribution
of fermionic higher spin particles, which have been neglected thus far in the 
literature. This necessarily introduces more than just a single additional 
fermionic superpartner, since the spectrum of massive, higher spin supermultiplets 
includes two propagating higher
spin bosons and two propagating higher spin fermions, which all contribute to 
the three point function. As an example we consider the half-integer superspin 
$\Ysf=s+1/2$ supermultiplet, which includes particles of spin values 
$j=s+1,~j=s+1/2,~j=s+1/2$ and $j=s$. We compute the curvature perturbation
3-point function for higher spin particle exchange and find that
the known $P_{s}(\cos \theta)$ angular dependence is accompanied by superpartner 
contributions that scale as $P_{s+1}(\cos \theta)$ and
$\sum_{m}P^{m}_{s} (\cos \theta)$, with $P_{s}$ and $P_{s} ^m$ defined as
the Legendre and Associated Legendre polynomials respectively. We also
compute the tensor-scalar-scalar 3-point function, and find a complicated
angular dependence as an integral over products of Legendre and associated Legendre polynomials.
}
\end{center}
$$~~$$
\vfill
%\noindent PACS: 11.30.Pb, 12.60.Jv\\
\vfill
\clearpage
%
%%%%%%%%%%%%%%%%%%%%%%%%%%%%%%%%%%%%%%%%%%%%%%
\section{Introduction}
\label{intro}
%%%%%%%%%%%%%%%%%%%%%%%%%%%%%%%%%%%%%%%%%%%%%%

Inflationary cosmology provides the initial conditions of standard cosmology, 
and a mechanism to explain the origin of the large scale structure of the 
universe. These initial conditions are manifest in the statistical properties of 
anisotropies in the cosmic microwave background (CMB) radiation, which in 
addition to being measured to an incredible precision \cite{Akrami:2018vks}, are 
well described by linearized cosmological perturbation theory. This latter fact 
means the statistical properties of the CMB are \emph{calculable}, and 
combined with the conservation of the primordial curvature perturbation on 
super-horizon scales 
\cite{Wands:2000dp,Weinberg:2003sw,Rigopoulos:2003ak,Lyth:2004gb,Langlois:2005qp,Assassi:2012et,Senatore:2012ya}, 
makes possible deductions as to the precise particle content of the very early 
universe.
 
In particular, while single-field slow-roll inflation predicts adiabatic, 
Gaussian, nearly-scale invariant perturbations,  interactions of the primordial 
curvature perturbation with new fields can generate deviations from Gaussianity, 
as encoded at lowest-order by the 3-point function $\langle \zeta \zeta \zeta 
\rangle$. Remarkably, this is sensitive even to fields that are heavier than the 
Hubble scale during inflation \cite{Chen:2009zp,Chen:2009we,Chen:2015lza}, and 
thus probes particles at energy scales far above those accessible by terrestrial 
colliders. The study of $\langle \zeta \zeta \zeta \rangle$ as a particle 
detector has been termed `Cosmological Collider Physics' \cite{AHM2015} (see 
also \cite{Lee:2016vti}), and gained significant momentum due to the realization 
that interactions with higher spin bosons, namely the exchange of a massive 
spin-$s$ boson, impart a characteristic angular dependence on the 
non-Gaussianity, $\langle \zeta(k_1) \zeta(k_2) \zeta(k_3) \rangle \propto 
P_{s}(\hat{k}_1 \cdot \hat{k}_3) + k_2 \leftrightarrow k_3$, with $P_{s}(\cos 
\theta)$ the degree-$s$ Legendre polynomial.

The study of higher spins has a long history
dating all the way back to the founding of relativistic field theory
\footnote{First paper by Majorana in 1932 \cite{s3-1} followed by Dirac, Pauli,
Fierz, Wigner and others.}. Since then, higher spins have gained fame and attention
in large part due to their role in string theory\footnote{For example, the UV softness 
of perturbative string scattering amplitudes originates from the freedom to exchange
higher spin particles. More recently, higher spin fields have played a role in constraining
the self-consistency of inflation in string theory \cite{Noumi:2019ohm,Lust:2019lmq}.}, as well as their use in exploring the holographic 
principle\footnote{All available, consistent, fully interacting higher spin theories
(such as Vasiliev's or CS in 3D) require an AdS background and a spin two state. 
Both of these requirements are ingredients of AdS/CFT correspondence.}.
Additionally, there is the old conjecture \cite{OG1,OG2,OG3,OG4} that physics beyond
Planckian energy scales will have higher symmetries emerging. From this point of view
the study of higher spins can be understood as an attempt to classify and realize
the various possibilities for these emerging symmetries. The study of manifestly
\emph{supersymmetric} higher spins is a natural extension of the above program, both because supersymmetry is a property of the 
underlying theory (as in the example of (super)string theory) and, more generally, because it is
compatible with the relevant structures
(like the symmetries of S-matrix \cite{s3-2}).

For these reasons we are interested in using irreducible representations of the 
supersymmetric extension of appropriate
spacetime symmetry groups which involve higher spin particles.
These irreps are classified and labeled by the eigenvalues of the
Casimir Operators. In $4D$ these are the mass ($m$) and the superspin ($\Ysf$) which 
takes either integer ($\Ysf=s$) or half integer values 
($\Ysf=s+1/2$)\footnote{For the purpose of this discussion we will ignore 
infinite sized representations that go under the name of continuous (super)spin 
representations \cite{Buchbinder:2019esz}.}. These representations include multiple representations of the 
non-supersymmetric spacetime symmetry group: For the massless case ($m=0$) a 
supermultiplet with superspin $\Ysf$ includes massless particles with spins 
$j=\Ysf+1/2$ and $j=\Ysf$, whereas for the massive case ($m\neq0$)
a supermultiplet with superspin $\Ysf$ includes massive particles with spins 
$j=\Ysf+1/2$, $j=\Ysf$, $j=\Ysf$ and $j=\Ysf-1/2$. This implies that supersymmetrizing the cosmological collider does not simply require adding a single higher spin fermion, but rather additional fields as well.

As a step towards combining this with inflationary cosmology, we consider an 
inflationary sector minimally coupled to a higher spin sector. The inflationary 
vacuum energy $H^2$ breaks supersymmetry, generating masses for the inflationary 
fermionic superpartners, while the higher spin sector, behaving as `spectator 
fields' which do not contribute to $H^2$ and hence do not contribute to 
supersymmetry breaking, retain their on-shell supersymmetry. Given the candidate 
bosonic interactions proposed in the literature \cite{AHM2015,Lee:2016vti}, the 
remnant on-shell supersymmetry of the higher spin sector uniquely determines the 
interactions of the higher spin fermions with the primordial curvature 
perturbation. From this one can compute the statistical properties of 
anisotropies in the CMB, and in this way, use the CMB as a detector for 
higher spin supersymmetry at the early universe's collider.

Each higher spin particle, as enumerated by the corresponding irreducible 
representation,  induces a contribution to the 3-point function $\langle \zeta 
\zeta \zeta \rangle$, i.e. a signal at the cosmological collider, and in this 
work we explicitly calculate the non-Gaussianity due to these contributions. Our primary result 
is the prediction of higher spin supersymmetry for the angular dependence of the 
non-Gaussianity: we find that the $P_s$ contributions to the non-Gaussianity 
come in a characteristic pattern. Namely, every $P_s$ contribution to the 
non-Gaussianity is accompanied by a $P_{s+1}$ contribution and a tower of 
associated Legendre polynomials $P_{s} ^m$.  The magnitudes, while Boltzmann 
suppressed, are related by supersymmetric considerations.

This paper is organized in the following way. 
In section 2 we build the effective field theory that will be the framework for 
our calculations. Section 3, gives a very elementary review of massive, higher 
superspin supermultiplets
focusing on the spectrum of propagating spin particles they include. Furthermore 
it demonstrates how to construct a supersymmetric extension of the previously
studied class of effective field theories.
In section 4, using these types of fermionic higher spin terms we consider an 
effective
interacting Lagrangian up to first order in the higher spin fields. The 
effective Lagrangian is then used
to calculate the contribution of higher spin fermions to the three point 
function $\langle \zeta\zeta\zeta \rangle$.
The last section 5, gives a summary of our results and a short discussion for 
future directions, including the tensor-scalar-scalar  3-point function $\langle 
\gamma \zeta \zeta \rangle$, which we compute for higher spin 
fermion exchange.

%%%%%%%%%%%%%%%%%%%%%%%%%%%%%%%%%%%%%%%%%%%%%%
\section{Setup in Effective Field Theory}
\label{sec2}
%%%%%%%%%%%%%%%%%%%%%%%%%%%%%%%%%%%%%%%%%%%%%%
%%

In this work we consider a tripartite marriage of $4D,\N=1$ supersymmetric higher spins
\cite{hss2,hss3,hss4,BKS,GKS96a,GKS,hss5,hss6,hss7,hss8,Sorokin,mhs1,mhs2,mhs3,mhs4} 
with the effective 
field theory (EFT) of inflation \cite{Cheung:2007st} and de Sitter supergravity 
\cite{Bergshoeff:2015tra,Kallosh:2015tea,Kallosh:2015sea}. We construct an effective field 
theory of a supersymmetric theory of higher spins in a quasi-de Sitter spacetime with 
spontaneously broken supersymmetry and spontaneously broken time-translation 
invariance. The goal of this construction is to minimally couple the higher spin sector
and the inflationary sector, in such a way that the on-shell 
supersymmetry of the higher spin fields is maintained, despite supersymmetry 
being broken by the inflationary vacuum energy.  The on-shell supersymmetry of 
the higher spin sector can then be used in conjunction with the effective field 
theory of inflation to dictate the couplings of higher spin fermions and bosons 
to primordial perturbations. This is distinct from the supersymmetric EFT of inflation \cite{Delacretaz:2016nhw} in that we do not focus on the gravity multiplet, but instead on the higher spin supermultiplets.

Our approach allows us to make progress despite not having a full theory of 
interacting higher spin de Sitter supergravity. While the setup may seem 
contrived, it bears some similarity with the interplay of supersymmetry and 
anomaly cancellation in string theory.

To appreciate this, one may recall that the interactions between effective field theory, supersymmetry and anomaly
cancellation are not as direct as one might imagine. In some cases a complete
superspace formulation or component-level supersymmetrization is known
such as in the example of the $4D,~\N=1$ WZNW-QCD action 
\cite{Gates:2000dq,Gates:2000gu,Gates:2000rp}.
In the development of Heterotic theory, 
anomaly cancellation required the addition of a new term in the action a la the 
famous Green-Schwartz mechanism \cite{Green:1984sg}, and this term required yet 
more terms (and years of calculation) in order to restore supersymmetry to the 
action\cite{Bergshoeff:1989de}. Similarly, in Type II/M theories the 
gravitational anomaly on D/M-branes induced by loops of chiral fermions is 
canceled via anomaly inflow by a higher-derivative correction to the bulk 
action \cite{Vafa:1995fj,Duff:1995wd,Witten:1996hc}, and despite the anomaly not 
playing any direct role in supersymmetry, as for Green-Schwarz the 
supersymmetrization of the anomaly-canceling terms requires yet more terms be 
added, the calculation of which requires a herculean level of technical skill 
and detail \cite{Peeters:2000qj}. The same issue applies to the 
$SL(2,\mathbb{Z})$ symmetry of type IIB: restoring the invariance naively broken 
by the corrections requires the careful consideration of D-instantons 
\cite{Green:1997tv} (and again these new terms must be supersymmetrized). 

In each of these cases, a seemingly complete theory is found to be anomalous, 
and cancellation of the anomalies requires new terms. The new terms should 
respect the symmetries of the action, and generically additional terms must be 
found to accomplish this task. However, much can be learned even \emph{without} 
a complete knowledge of all terms in the theory. For example, in the interim 
period between  \cite{Vafa:1995fj,Duff:1995wd,Witten:1996hc} and 
\cite{Peeters:2000qj}, the AdS/CFT correspondence was discovered 
\cite{Maldacena:1997re}. Another example of this is of course the field of 
String Cosmology \cite{Baumann:2014nda,Alexander:2001ks}, which makes no recourse to the precise 
manner in which $SL(2,\mathbb{Z})$ symmetry is maintained. With all this in 
mind, we construct an effective theory along the lines discussed above.

This approach can be illustrated with simple examples involving chiral 
superfields in ${ \cal N}=1$ supersymmetry. The first non-trivial step is the 
`sequestering' of supersymmetry breaking to the inflationary sector, analogous 
to the Randall-Sundrum scenario \cite{Randall:1998uk}. This can be done in a 
number of ways; one simple example is to take guidance from  
\cite{McDonough:2016der,Kallosh:2017wnt} and allow for a non-minimal Kahler 
potential, as in
\be
W = M X  \;\; ,\;\;
 K = X \bar{X} e^{Y \bar{Y}}+  Y \bar{Y}  .
\ee
This exhibits a vacuum at $X=\bar{X}=Y=\bar{Y}=0$, wherein supersymmetry is 
broken by $X$, $D_{X}W=M$. The scalar potential evaluated for $X=\bar{X}=0$ is 
given by a constant $V =M^2 $, leaving the scalar component of $Y$ massless: 
$m^2 _{Y} \equiv \partial_{Y \bar{Y}} V=0$ \footnote{For this simple example, 
also the scalar component of $X$ is massless, but it can be made massive via an 
addition to the Kahler potential $\delta K = (X \bar{X})^2 / \Lambda$. The 
fermionic component of $X$ has mass set by $M$.}.  Similarly, the fermion 
component of $Y$ remains massless since $D_{Y}W = 0 $. Thus the breaking of 
supersymmetry is not communicated to the on-shell mass spectra of $Y$, and the 
$Y$ superfield retains on-shell $\mathcal{N}=1$ supersymmetry.

The utility of this approach is the enumeration of interactions and the 
tree-level couplings, since despite the sequestering of supersymmetry breaking, there are 
interactions between $X$ and $Y$, which communicate the SUSY breaking at 
loop-level. Expanding $X\rightarrow \delta x$ and $Y \rightarrow \delta y$, 
$\delta x,\delta y \in \mathbb{R}$, one finds the interactions between scalar 
components,
\be
\mathcal{L}_{int}= \delta y^2 (\partial \delta x)^2 + M^2 \delta y^2 \, \delta x^4 + ... ,
\ee
where the $...$ are higher order terms. Similarly, there are  interactions 
between the fermionic components of $X$ and the fermionic components of $Y$, and 
these two sets of interactions will communicate the SUSY breaking to $Y$. The 
structure of these interactions is governed by the underlying supersymmetry, 
which is spontaneously broken by $X$, and this structure dictates the effect 
that $\delta y$ interactions can have on $\delta x$ correlators.

In our setup, the higher spin sector is analogous to $Y$ while the inflationary 
sector is analogous to $X$.  It is the above sense in which the HS sector in our 
setup has on-shell supersymmetry. This can be used to enumerate the interactions 
and estimate the amplitude of correlation functions. However, this is not the 
full story: The next puzzle piece is the embedding of supersymmetry and 
supergravity into cosmological spacetimes. 

This can be done via the framework of {\it de Sitter supergravity} 
\cite{Bergshoeff:2015tra,Kallosh:2015tea,Kallosh:2015sea}. This theory describes 
the spontaneous breaking of supersymmetry with no field content other then the 
gravity multiplet and the goldstino of supersymmetry breaking. The latter can be 
expressed as a chiral superfield, $S$, satisfying a constraint equation,
\be
S^2 = 0 .
\ee
This constraint removes the scalar degree of freedom from $S$, leaving the 
fermionic component as the only propagating degree of freedom. The most general 
superpotential is given by,
\be
W = W_0 + M S ,
\ee
since any additional terms involving $S$ vanish by the nilpotency constraint. 
Supersymmetry is broken by $D_S W=M$, and the resulting scalar potential, for a 
minimal Kahler potential $K=S \bar{S}$, is a cosmological constant given by
\be
\Lambda \equiv V  = M^2 - 3 W_0^2 ,
\ee
which is positive for $M> \sqrt{3} W_0$, giving a de Sitter spacetime.

Any additional matter sectors in de Sitter supergravity can easily be 
sequestered from the breaking of supersymmetry. For example, endowing $S$ with a 
non-trivial Kahler geometry \cite{McDonough:2016der,Kallosh:2017wnt} and taking 
$W_0=0$,
\be
W = M S \;\; ,\;\; K =  e^{T \bar{T}}S \bar{S} + T \bar{T} \;\; ,\;\; S^2=0,
\ee
supersymmetry-breaking is purely in the $S$-direction provided that $D_{T}W=0$ 
in vacuum, which is guaranteed to be the case since $D_{T}W \propto S =0$, 
leaving the fermionic component of $T$ massless. Meanwhile, SUSY is broken by 
$S$, $D_S W = M$, and the potential is a constant vacuum energy $V=\Lambda=M$, 
leaving the scalar components of $T$ massless. Thus again, $T$ retains on-shell 
supersymmetry.

To connect this with observational cosmology, and anisotropies in the cosmic 
microwave background radiation, it is necessary to consider fluctuations. 
Inflation models can be constructed in de Sitter supergravity along the lines of 
\cite{McDonough:2016der,Kallosh:2017wnt,Ferrara:2014kva,Kallosh:2014via}. 
Consider a superfield $\Phi$ with the real part of the scalar component of 
$\Phi$ identified as the inflaton $\varphi$.  The fluctuations of $\varphi$ in 
spatially flat gauge are related to the curvature perturbation on uniform 
density hypersurfaces $\zeta$ in uniform field gauge via  
\cite{Brandenberger:2003vk}
\be
\zeta = \frac{H}{\dot{\varphi}} \delta \varphi = \frac{1}{\sqrt{2 \epsilon}m_{pl}}\delta \varphi ,
\ee
where $\epsilon \equiv - \dot{H}/H^2$ is 
the inflationary slow-roll parameter. This defines the primordial power 
spectrum, in dimensionless form,
\be
\Delta_{\zeta} ^2 \equiv \frac{k^3}{2 \pi^2} |\zeta_k|^2 ,
\ee
where $\zeta_k$ is a Fourier mode of the field $\zeta(x,t)$.

The curvature perturbation $\zeta$ can in turn be related to the Goldstone boson 
of spontaneously broken time-translation invariance, using the machinery of the 
effective field theory of inflation  \cite{Cheung:2007st}. This starts from the 
realization that the time-dependence of the inflaton $\varphi(t)$ breaks the time 
diffeomorphisms. The Goldstone boson can be included in the theory by a 
redefinition of the time-coordinate,
\be
t \rightarrow t - \pi(t,x) ,
\ee
with $\pi(t,x)$ the Goldstone boson. This induces a field fluctuation,
\be
\varphi(t) \rightarrow \varphi(t) - \dot{\varphi} \pi(t,x)
\ee
and thus corresponds to a curvature perturbation,
\be
\zeta = - H \pi .
\ee
This also generates a fluctuation to the $00$ component of the metric,
\be
\delta g^{00} = -2 \dot{\pi} .
\ee
The interactions of $\pi$, and hence $\zeta$, are dictated by the symmetry 
structure of the action, which is broken to invariance under spatial 
rotations. This allows $\delta g^{00}$ to appear explicitly in the action, while 
$\delta g^{ij}$ can only appear with all indices contracted.

Similarly, the interactions of $\pi$, and hence $\zeta$, with any additional fields are 
dictated by symmetry considerations. For fields with arbitrary spin, 
$\sigma_{\mu_1 ... \mu_s}$, this leads to effective interactions of the form 
\cite{AHM2015,Lee:2016vti}
\be
\label{HSbosonints}
\mathcal{L}_{int} \supset  \frac{\lambda_s}{\Lambda^{s-3}}\partial_{i_1}
... \partial_{i_s} \zeta \sigma ^{i_1 ... i_s}+ \frac{g_s}
{\Lambda^{s-2}} \dot{\zeta} \partial_{i_1} ... \partial_{i_s} \zeta  \sigma ^{i_1 ... i_s}~,
\ee
where indices $i$ runs over spatial directions: $1,2,3$, and $\Lambda$ is a UV scale. These terms descend from 
higher-dimension operators built out of the metric and its derivatives, and have 
coupling constants that are \emph{a priori} free parameters of the effective 
field theory. For example, in the spin-2 case, these terms arise from a coupling of a spin-2 field $\sigma$  to the extrinsic curvature, $\sqrt{-g}K^{\mu \nu} 
\sigma_{\mu \nu}$.  The first term in \eqref{HSbosonints} descends from $\delta K^{\mu \nu} 
\sigma_{\mu \nu}$  while the second term arises from including the metric perturbation $\delta g^{00} \delta K^{\mu \nu} 
\sigma_{\mu \nu}$ \cite{Lee:2016vti}. There can also be additional terms at 
higher order in $\pi$ and $\sigma$ and terms with different distribution 
of derivatives (up to total derivatives).

We now arrive back at the tripartite marriage: We wish to connect the 
higher spin interactions in a quasi-dS space  to supersymmetry. To do this, one 
\emph{could} simply operate along effective field theory lines, and introduce 
interactions consistent with unbroken spatial rotations. However, an interesting 
possibility is to consider what we can learn from higher spin supersymmetry, and 
use this as guidance in constructing our effective field theory describing the 
interactions of the higher spin fermions. Towards this end, we now develop the 
machinery of higher spin supersymmetry.

%%%%%%%%%%%%%%%%%%%%%%%%%%%%%%%%%%%%%%%%%%%%%%
\section{Supersymmetric Higher Spins}
\label{shs}
%%%%%%%%%%%%%%%%%%%%%%%%%%%%%%%%%%%%%%%%%%%%%%

The first Lagrangian description of supersymmetric, massless, higher spins in 
$4D$ Minkowski space was done in \cite{hss1,V}, using components with on-shell
supersymmetry. A natural approach to the off-shell formulation is to use the
superspace and superfield methods (see e.g.\cite{GGRS,BK}). A superfield
description of free supersymmetric massless, higher spin theories was presented 
for the first time in \cite{hss2,hss3,hss4} for both Minkowski and AdS spaces. This 
approach has been further explored in \cite{BKS,GKS96a,GKS,hss5}.
Later studies of free supersymmetric, massless higher spin supermultiplets include
\cite{hss6,hss7,hss8,Sorokin}.

On the other hand, the Lagrangian description of 4D massive
supersymmetric spins for arbitrary values of spin is only known in the component formulation 
with on-shell supersymmetry \cite{mhs1,mhs2}, whereas 
the off-shell, superspace description has been developed up to
superspin $\Ysf=3/2$ supermultiplet \cite{mhs3,mhs4}. 
Nevertheless, independently of what the proper Lagrangian description is,
we know that there are two types of such irreps (\emph{i}) the integer superspin $\Ysf=s$
supermultiplets and (\emph{ii}) the half integer superspin $\Ysf=s+1/2$
supermultiplets. Moreover we know that on-shell they describe two bosonic and
two fermionic massive higher spin particles with spin values
$j=\Ysf+1/2$,~$j=\Ysf$,~$j=\Ysf$ and $j=\Ysf-1/2$. The half integer superspin
$\Ysf=s+1/2$ supermultiplet, consisting of components
\be
\label{Yhalfint}
\Ysf=s+1/2 : \;\;\; (s+1,~s+1/2,~s+1/2,~s)
\ee
and its on-shell, superspace description is given in terms of a real, bosonic superfield 
$H_{\a(s)\ad(s)}$\footnote{The notation $\a(s)$ signifies a string of $s$
undotted indices $\a_1\a_2...\a_s$ which are symmetrized. This type of indices are
the spinorial indices of a Weyl spinor of one chirality and take values 1 and 2 in $4D$.
Similarly for $\ad(s)$, where $\ad$ are the spinorial indices of the opposite chirality Weyl spinor
and take also two values $\dot{1}$ and $\dot{2}$ in $4D$.} with the following on-shell conditions:
\bea
\D^{\a_s}H_{\a(s)\ad(s)}=0~,~
\Box H_{\a(s)\ad(s)}=m^2 H_{\a(s)\ad(s)}~ ,\n
\eea
where $D^{\a_s}$ is the superspace covariant derivative. Alternatively,
the massive, integer $\Ysf=s$ superspin supermultiplet, comprised of components,
\be
\label{Yint}
\Ysf=s : \;\;\; (~s+1/2,~s,~s,~s-1/2) 
 \ee 
has an on-shell superspace description in terms of a fermionic superfield $\Psi_{\a(s)\ad(s-1)}$ with the on-shell equations:
\bea
\D^{\a_s}\Psi_{\a(s)\ad(s-1)}=0~,~
\Dd^{\ad_{s-1}}\Psi_{\a(s)\ad(s-1)}=0~,~
i\pa_{\a_s}{}^{\ad_s}\bar{\Psi}_{\a(s-1)\ad(s)}+m\Psi_{\a(s)\ad(s-1)}=0~.\n
\eea

We start with the assumption that the higher 
spin sector respects supersymmetry and therefore can be organized
into higher spin supermultiplets. This is extremely useful because 
supersymmetry will guide us to the introduction of higher spin fermions
which have been neglected so far. Once their contribution is better understood, 
one may choose to drop the assumption of supersymmetry and study these 
fermionic contributions independently.

The strategy for finding the fermionic higher spin contributions is: (\emph{a}) Start with the
family of effective actions that lead to (\ref{HSbosonints}) after breaking the time translation invariance,
and elevate them to superspace. This will automatically introduce all fermionic partners. (\emph{b}) We project
back down to a component description to reveal the interactions of the higher spin fermions. (\emph{c}) Finally, we break supersymmetry appropriately in the
inflaton sector.

The first step is to embed the bosonic, massive spin $s$ particle in
a massive higher spin supermultiplet described by some higher spin superfield.
As we have seen, there are two ways of doing that, we can either
use the integer or the half-integer superspin supermultiplet, with components \eqref{Yint} and \eqref{Yhalfint} respectively. 
For concreteness, we make the latter choice
($\Ysf=s+1/2$) which means that our spin $s$ particle will be accompanied
by one bosonic higher spin particle $j=s+1$ and two more fermionic
higher spin particles $j=s+1/2$. In this choice the highest propagating
spin is $s+1$. Similarly we embed the scalar curvature perturbation field in a scalar supermultiplet,
which will of course introduce its fermionic superpartner, which we refer to as the inflatino
\footnote{In general the fermionic partner of the curvature perturbation field can be identified as
a linear combination of the inflatino and other fermions in the theory.}. A simple choice
to describe such a scalar supermultiplet is to use a chiral superfield $\Phi$.

Secondly, using superspace, we write quadratic and cubic interaction terms,
between $H_{\a(s)\ad(s)}$ and $\Phi$ which are linear in the higher spin superfield.
The family of such superspace effective Lagrangians takes the form:
\bea
\mathcal{L}=H^{\a(s)\ad(s)}\mathcal{I}_{\a(s)\ad(s)}~+~H^{\a(s)\ad(s)}\J_{\a(s)\ad(s)}\n
\label{esft}
\eea
where $\mathcal{I}_{\a(s)\ad(s)}$ is linear in $\Phi$ and generates
the quadratic interactions part, whereas $\J_{\a(s)\ad(s)}$
is quadratic in $\Phi$ and generates the cubic
\footnote{The reason why we are considering massive higher spin supermultiplets is a consequence
of the Higuchi bound \cite{Higuchi:1986py,Deser:2001us} plus the possibility of higher order mass-like
interaction terms for the higher spin superfields, which we do not consider in this work. If that
was not the case one should take into account the gauge symmetry of the higher spin (super)fields.
The result of that would be that, the spectrum of the half integer supermultiplet will collapse from
($s+1,~s+1/2,~s+1/2,~s$) to ($s+1,~s+1/2$) and more importantly the generator of cubic interactions
$\J_{\a(s)\ad(s)}$ in (\ref{esft}) will become conserved higher spin supercurrents. Such supercurrents have been found
in \cite{sc1,sc2,sc7,sc6,sc3,sc4,sc8,sc5,sc9} and their structure is consistent with (\ref{J}).
}
part of the interactions.
The most general ansatzes for $\mathcal{I}_{\a(s)\ad(s)}$ and $\J_{\a(s)\ad(s)}$ are,
\bea
&&\hspace{-5ex}\mathcal{I}_{\a(s)\ad(s)}=\pa^{(s)}(b~\Phi+b^*~\bar{\Phi}) ,
\eea
and
\bea
&&  \hspace{-8ex} \J_{\a(s)\ad(s)}=
\vphantom{\frac12}
\sum_{p=0}^{s}\left\{\vphantom{\frac12}
d_{p}~\pa^{(p)}\Phi~\pa^{(s-p)}\bar{\Phi}+
f_{p}~\pa^{(p)}\D\Phi~\pa^{(s-p-1)}\Dd\bar{\Phi}
+g_{p}~\pa^{(p)}\Phi~\pa^{(s-p)}\Phi+g^{*}_{p}~\pa^{(p)}\bar{\Phi}~\pa^{(s-p)}\bar{\Phi}\right\}\label{J} .
\eea

Using this as a starting point, one can project the 
superspace Lagrangian to components and find the corresponding
field theory (see \cite{GGRS,BK}
and detailed examples can be found in \cite{hss5,sc2}). The result will include the entire spectrum of fields of the 
supersymmetric theory. In addition to the propagating spins,
this includes the set of auxiliary fields required by supersymmetry in order to 
balance the bosonic and fermionic degrees of freedom and also make the symmetry manifest.
However, these auxiliary fields do not have any dynamics and can be integrated out.
By doing so, we obtain an effective theory with on-shell supersymmetry
which includes two copies of the previously discussed bosonic higher spin interactions.
That is because there are two
higher spin bosons, one with spin $s$ and one with spin $s+1$. Additionally, we obtain terms
that depend on the higher spin fermions ($\psi_{\a(s+1)\ad(s)}$,~$\xi_{\a(s+1)\ad(s)}$)
and the `inflatino' ($\chi_{\a}$). The interactions are given by,
\bea
&&\hspace{-7.2ex}\mathcal{L}_\text{Bosonic}=h^{\a(s+1)\ad(s+1)}
\left[\vphantom{\frac12}\tfrac{\lambda_{s+1}}{\Lambda^{s-2}}~\pa^{(s+1)}\zeta
+\sum_{p=0}^{s+1}\tfrac{\kappa_{p}^{s+1}}{\Lambda^{s-2}}~\pa^{(p)}\zeta~\pa^{(s+1-p)}\zeta
+...\right]\\
&&\hspace{2ex}+h^{\a(s)\ad(s)}\left[\vphantom{\frac12}\tfrac{\lambda_{s}}{\Lambda^{s-3}}~\pa^{(s)}\zeta
+\sum_{p=0}^{s}\tfrac{\kappa_{p}^{s}}{\Lambda^{s-3}}~\pa^{(p)}\zeta~\pa^{(s-p)}\zeta
+...\right]~ \label{HSbosoniccouplings}\nonumber\\
&&\hspace{-7.2ex}\mathcal{L}_\text{Fermionic}=\psi^{\a(s+1)\ad(s)}
\left[\vphantom{\frac12}\tfrac{r_{s}}{\Lambda^{s-1}}~\pa^{(s)}\chi_{\a}
+\tfrac{\lambda_{s}}{\Lambda^{s-1}}~\chi_{\a}\pa^{(s)}\zeta
+\sum_{p=1}^{s}\tfrac{v_{p}}{\Lambda^{s-1}}~\pa^{(p)}\chi~\pa^{(s-p)}\zeta
+...\right]+h.c.\label{HSfermioncouplings}\\
&&\hspace{4ex}+\xi^{\a(s+1)\ad(s)}
\left[\vphantom{\frac12}\tfrac{t_{s}}{\Lambda^{s-1}}~\pa^{(s)}\chi_{\a}
+\tfrac{\lambda_{s}}{\Lambda^{s-1}}~\chi_{\a}\pa^{(s)}\zeta
+\sum_{p=1}^{s}\tfrac{w_{p}}{\Lambda^{s-1}}~\pa^{(p)}\chi~\pa^{(s-p)}\zeta
+...\right]+h.c.\nonumber
\eea
where $...$ indicates additional and higher-order terms.

As discussed in section \ref{sec2}, on-shell supersymmetry should be preserved 
only in the higher spin sector, and not in inflaton sector which breaks
supersymmetry with the inflationary vacuum energy $H^2$. In our effective 
Lagrangian this information can be entered
by hand 
by removing any correlation between the coupling constants of the inflatino and 
those of the inflaton.
For example we can assign the inflatino $\chi$ a mass $m_{\chi} \gtrsim H$. 
Typically one would then {\it integrate out} the inflatino, 
thereby eliminating all the contributions at linear order in the higher spin 
fermions \eqref{HSfermioncouplings}. However it is the inclusion 
of such heavy fields that we are explicitly after in this work. Indeed, the 
higher spin fields themselves have mass $m > \sqrt{s (s-1)}H$ by 
the Higuchi bound \cite{Higuchi:1986py,Deser:2001us}.
The breaking of supersymmetry will also induce differing loop corrections to the 
on-shell couplings of $\zeta$ to bosonic 
\eqref{HSbosoniccouplings} and fermionic \eqref{HSfermioncouplings} higher spin 
fields. Depending on the precise details of the model, there 
may also be classical corrections to these parameters, for example, from a 
quartic interaction involving an additional scalar field that 
gains a VEV in the SUSY-breaking vacuum. For our analysis, we assume for 
simplicity that there are no such classical corrections, and that 
these couplings are equal at tree level. This does not alter the analysis in any 
way other then the overall prefactor of the result.

To make contact with the framework of effective field theories within we have to work, as presented in section \ref{sec2}, we must break
the time translations part of the Poincar\'e group and write interaction terms which include
fermionic higher spin particles up to linear order.
From equation \eqref{HSfermioncouplings}, and taking into
account contributions coming from the $\sqrt{-g}$ part of the action, one has to consider the following fermionic interaction Lagrangian for a spin-$s+1/2$ field:
\be
\label{LHS1fa}
\mathcal{L}  \supset 
\frac{\lambda_s}{\Lambda^{s-1}}\partial_{i_1 ... i_s} 
\zeta \bar{\chi} \psi ^{i_1 ... i_s}
+ \frac{g_s}{\Lambda^{s}}\dot{ \zeta} 
\partial_{i_1 ... i_s} \zeta \bar{\chi} \psi ^{i_1 ... i_s}+\frac{\kappa_s}{\Lambda^{s}}\dot{\zeta}\partial_{i_1 ... i_s} 
\bar{\chi} \psi ^{i_1 ... i_s}+c.c.,
\ee
where the coupling to $\dot{\zeta}$ enters from the metric perturbation $\delta g^{00}$, as in equation \eqref{HSbosonints}. The full fermionic interaction Lagrangian is composed of copies of \eqref{LHS1fa} for the appropriate fermions in the supermultiplet. Armed with this, we can now return to the cosmological collider.

%%%%%%%%%%%%%%%%%%%%%%%%%%%%%%%%%%%%%%%%%%%%%%
\section{Higher Spin Supersymmetry at the Cosmological Collider}
\label{sec4}
%%%%%%%%%%%%%%%%%%%%%%%%%%%%%%%%%%%%%%%%%%%%%%

%   
The statistical correlations of temperature fluctuations in the cosmic microwave 
background descend from the initial conditions prepared for it by inflation. 
This can be computed via the Schwinger-Keldysh formalism, colloquially called 
the `In-In' formalism, in which the choice of integration contour allows for 
ignorance as to the future evolution of the universe. Introduced to cosmology in 
\cite{Maldacena:2002vr}, there are now many excellent reviews on this topic, see 
e.g. \cite{Chen:2010xka,Adshead:2009cb,Wang_2014}, and see 
\cite{altland_simons_2010} for a textbook treatment of the field theory aspects.

Correlation functions can be computed in this framework by splitting the 
Hamiltonian into a free Hamiltonian $H_0$ and interaction Hamiltonian 
$H_{int}$. From this, one can define the interaction picture fields as having 
propagator determined solely by $H_0$, and correlation functions of operators 
built from the full fields can be computed as contractions of the interaction 
picture fields with the interaction Hamiltonian.

More precisely, the expectation value of an operator $W$ is given by,
\be
\label{Winin}
\langle W(t) \rangle = \langle \left[  \bar{\rm T} e^{ i \int _{-\infty 
(1-i\epsilon)} ^t H_{int} ^{I} (t') {\rm d}t'}\right] W^I (t)  \left[  {\rm T} 
e^{- i \int _{-\infty (1+i\epsilon)} ^t H_{int} ^{I} (t'') {\rm d}t''}\right]  
\rangle,
\ee
where $H^I _{int}$ and $W^I$ are the interaction Hamiltonian and the operator 
$W$ built out of interaction picture fields.  The simplest quantity one can 
compute from this is the expectation value given a single insertion of the 
interaction Hamiltonian. In that case, the above expression reduces to 
\be
\langle W(t) \rangle = - 2 \, {\rm Re} \; \langle \; i \, W^I(t) \int _{-\infty 
(1-i\epsilon)} ^t H_{int} ^{I} (t') {\rm d}t'  \;\rangle + ... ,
\ee
where the $...$ corresponds to additional insertions of $H_{int}$. 

For the case 
of the curvature perturbation 3-point function, $W= \zeta^3$, this picks out the 
intrinsic non-Gaussianity, first computed in \cite{Maldacena:2002vr}. This is 
the 3-point function induced by the self-interactions of $\zeta$ in an 
inflationary background. The result is given by
 \bea
&&   \langle \zeta(k_1,0)  \zeta(k_2,0) \zeta(k_3,0) \rangle  =  (2 \pi)^3 
\delta({k}_1 + { k}_2 + { k}_3) \frac{H^4}{m_{pl} ^4} \frac{1}{(k_1 
k_2 k_3)^3} \frac{1}{4 \epsilon^2} \nonumber \\
 && \cdot \left[ \frac{\eta}{8} \sum k_i ^3 + \frac{\epsilon}{8} \left( - \sum 
k_i^3 + \displaystyle \sum_{i \neq j} k_i k_j ^2 + \frac{8}{k_1 + k_2+ k_3} 
\sum_{i > j} k_i ^2 k_j ^2\right)\right] .
 \eea
This is typically expressed in the limit that one of the momenta is much smaller 
than the other two, in what is referred to as `squeezed limit'. The result then 
takes a simplified form
\be
\lim _{k_1 \rightarrow 0}   \langle \zeta(k_1,0)  \zeta(k_2,0) \zeta(k_3,0) 
\rangle = (2 \pi)^3 f_{NL} \delta({ k}_1 + { k}_2 + {k}_3) \frac{H^4}{16 \epsilon^2 m_{pl} 
^4} \frac{1}{(k_1 k_2 k_3)^3}  \sum k_i ^3 ,
\ee
which corresponds to `local shape' non-Gaussianity \cite{Komatsu:2001rj}  with 
amplitude $f_{NL}$ given by
\be
f_{NL}  = \frac{\eta}{2} + \epsilon.
\ee
The inflationary slow-roll conditions $\epsilon,\eta \ll 1$ thus imply the 
intrinsic non-Gaussianity in single-field slow-roll inflation is extremely 
small, $f_{NL}  \ll 1$.

Additional insertions of $H_{int}$ capture the effect of particle exchange.  Given the slow-roll suppression of the intrinsic non-Gaussianity, this can 
easily be the dominant effect. It is in this sense that CMB non-Gaussianity is a 
particle detector, with inflation as the cosmological collider. 

In this work we 
are concerned with the non-Gaussianity induced via the exchange of a higher spin 
particle, as described by the insertion of two interaction Hamiltonians. This is captured by the quadratic terms in the expansion of  \eqref{Winin}, see e.g.~\cite{Baumann:2017jvh},
\bea 
&& \langle W(t))\rangle = \int {\rm d}t'\int {\rm d}t'' \langle H^I(t') W(t) H^I(t'')\rangle  - 2\, {\rm Re} \int {\rm d}t' \int {\rm d}t'' \langle W(t)H^I(t')H^I(t'') \rangle ,
\label{eq:threept}
\eea 
with appropriate $i \epsilon$ prescriptions in the integrations.

%%%%%%%%%%%%%%%%%%%%%%%%%%%%%%%%%%%%%%%%%%%%%%

\subsection{Effective Action and Relevant Interactions}

%%%%%%%%%%%%%%%%%%%%%%%%%%%%%%%%%%%%%%%%%%%%%%

We consider an effective action describing the interactions of  a scalar 
$\zeta$, a massive spin-$1/2$ field $\chi$, and the propagating component fields 
of the massive, half-integer superspin $\Ysf=s+1/2$ supermultiplet, as they have 
been discussed previously. We consider our action as an expansion in higher spin 
fields, keeping up to linear order terms. We consider
\be
\mathcal{L} = \mathcal{L}_{\zeta} +  \mathcal{L}_{\chi}+  \mathcal{L}_{hs} + 
\mathcal{O}(hs^2)...
\ee

\begin{figure}[t]
  \centering
  \begin{subfigure}[t]{0.45\linewidth}
    \includegraphics[width=\textwidth,height=2in]{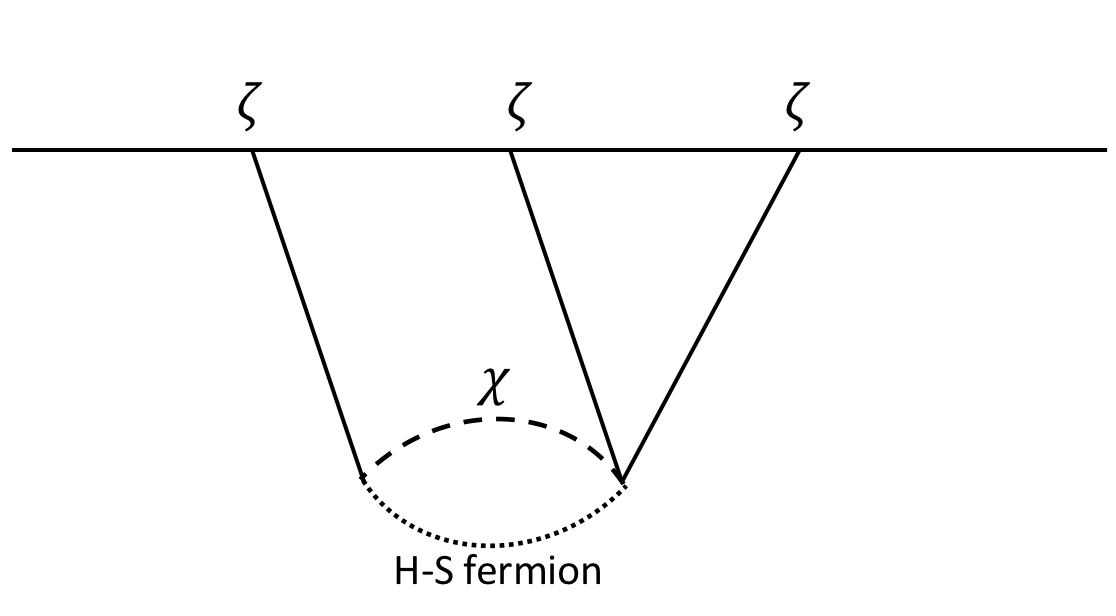}
    \caption{\label{fig:fig1} Exchange of a higher spin Fermion}
  \end{subfigure}%
  \begin{subfigure}[t]{0.45\linewidth}
    \includegraphics[width=\textwidth,height=2in]{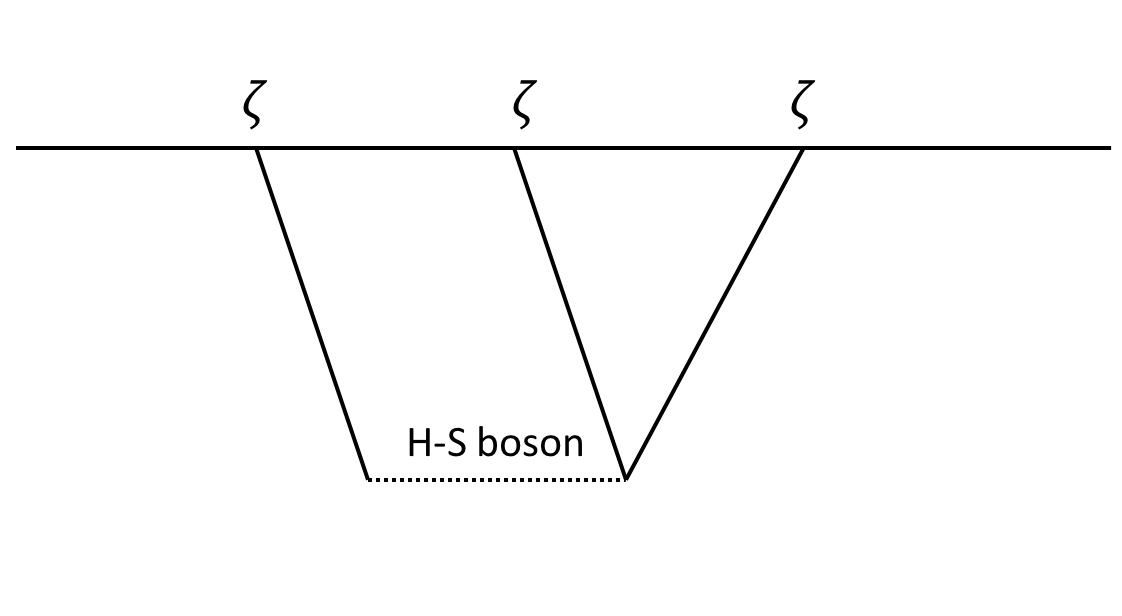}
    \caption{\label{fig:fig2} Exchange of a higher spin boson}
  \end{subfigure}
  \caption{In-In formalism Feynman Diagrams for exchange of a single higher spin particle.}
  \label{fig:HSdiagrams}
\end{figure}

In this work we are particularly interested in the impact of the fermionic higher 
spin particles, which has thus far been left unstudied. Motivated from the 
discussion in section 3 and (\ref{LHS1fa}), we take the Lagrangian of one fermionic spin-$s+1/2$ 
particle  $\psi$ interacting with a dimensionless scalar $\zeta$ and the 
spin-1/2 particle $\chi$, 
\be
\label{LHS1f}
\mathcal{L}  \supset 
\frac{\lambda_s}{\Lambda^{s-1}}\partial_{i_1 ... i_s} 
\zeta \bar{\chi} \psi ^{i_1 ... i_s}
+ \frac{g_s}{\Lambda^{s}}\dot{ \zeta} 
\partial_{i_1 ... i_s} \zeta \bar{\chi} \psi ^{i_1 ... i_s}+c.c.,
\ee
where $\lambda_s$ and $g_s$ are dimensionless coupling constants, $\Lambda$ is a 
UV cutoff, and fermionic indices are contracted between $\bar{\chi}$ and $\psi$. For simplicity we have taken $\kappa_s=0$ in  (\ref{LHS1fa}), which leads to the same angular dependence for $\langle \zeta \zeta \zeta \rangle$ as the two terms above.
The relevant Feynman diagrams, or rather their equivalent in the 
Schwinger-Keldysh (``in-in'' formalism) are shown in Figure 
\ref{fig:HSdiagrams}.

%%%%%%%%%%%%%%%%%%%%%%%%%%%%%%%%%%%%%%%%%%%%%%
\subsection{Higher Spin Fields in de Sitter Space}
\label{sec:HSdS}
%%%%%%%%%%%%%%%%%%%%%%%%%%%%%%%%%%%%%%%%%%%%%%

To evaluate the three-point function in the Schwinger-Keldysh formalism, we 
first must have expressions for the free fields in de Sitter space.

To begin with, a scalar field $\phi$ is quantized in curved space as,
\be
\phi(x,t) = \int \frac{\mathrm{d}^3k}{(2 \pi)^3} \phi_k a_k e^{ik\cdot x} + \;\; 
h.c. \; .
\ee
The coefficients $\phi_k (k,t)$ are referred to as ``mode functions''. 
The field $\phi(x)$ and mode functions $\phi_k$ are related to two-point 
correlation functions as follows. The position-space two point function of a 
free-field $\phi$ is given by,
\be
\langle \phi(x) \phi(y) \rangle =  \int \frac{\mathrm{d}^3 k}{(2 \pi)^3}   
\langle \phi_k \phi_k \rangle  \, e^{i k( x -y)}  =  \int \mathrm{d} \log k  \, 
\frac{k^3}{2 \pi^2} |\phi_{k}|^2  \, e^{i k( x -y)}  \equiv \int \mathrm{d} \log 
k  \, \Delta^2 _{\phi}(k)  \, e^{i k( x -y)}  .
\ee
The  last equality defines the dimensionless \emph{power spectrum}, $\Delta^2 
_{\phi} (k) \equiv \frac{k^3}{2 \pi^2} |\phi_{k}|^2$.
A special case of the above is a \emph{scale-invariant spectrum}. In this case, 
$\Delta ^2_{\phi} (k)$ is a constant,
which owes its name to the implied scaling symmetry of the two-point function, 
$\langle \phi(x) \phi(y) \rangle = \langle \phi(\lambda x ) \varphi (\lambda y) 
\rangle $. During inflation, this scaling symmetry of a massless scalar has its 
origins in the dilatation symmetry at late times in de Sitter space.

An important example is the curvature perturbation on uniform density 
hypersurfaces, $\zeta$. This has mode function given by,
\be
\label{zetak}
\zeta_k \simeq \frac{H}{m_{pl}\sqrt{4 \epsilon k^3}} (1 - i k \eta) e^{i k 
\eta},
\ee
where $\epsilon = - \dot{H}/H^2 \ll 1$ is the 
inflationary slow-roll parameter, and $\eta$ is conformal time ${  \rm d}\eta \equiv  a^{-1} { \rm d}t$, which in de Sitter space is given by $\eta = - 1/(aH)$. This solution is found by explicitly solving 
the Klein-Gordon equation in de Sitter space. Inflation is a small deviation 
from de Sitter space, which converts the scaling with $k$ to $k^{-3/2 + 
(n_s-1)/2}$, where $n_s$ defines the spectral index of the power spectrum,
 \be
 \label{Deltazeta}
 \Delta_{\zeta}^2 \equiv \frac{k^3}{2 \pi^2} |\zeta_k|^2 \propto k^{n_s -1} .
 \ee
Another important example is massless spin-2, e.g. the graviton. Expanded in 
helicity states $\lambda=\pm 2$, the mode functions are given by \cite{Baumann:2014nda}
\be
\gamma ^{\lambda} _k = \frac{\sqrt{2} H}{m_{pl}} \frac{1}{\sqrt{k^3}} (1 + i k \eta) 
e^{ -i k \eta} ,
\ee
which, importantly, differs from the curvature perturbation \eqref{zetak} in part by an factor of $1/\sqrt{\epsilon}$.  The power spectrum of primordial gravitational waves on large scales $k\eta\rightarrow 0$ is then given by \cite{Baumann:2014nda},
\be
\label{Deltagamma}
\Delta^2 _{\gamma}  \equiv \sum_{\lambda} \frac{k^3}{2 \pi^2} |\gamma ^{\lambda} _k| = \frac{2}{\pi^2}\frac{H^2}{ m_{pl} ^2},
\ee
which is a direct probe of the energy scale of inflation \cite{Abazajian:2016yjj}. The ratio of the tensor power spectrum \eqref{Deltagamma} to scalar power spectrum, termed the `tensor-to-scalar ratio', is given by
\be
r \equiv  \frac{\Delta^2_{\gamma}}{\Delta^2_\zeta} = 16 \epsilon,
\ee
where again $\epsilon \ll 1$ is the inflationary slow-roll parameter.

In contrast with these two examples, for massive particles the  two-point function and hence mode functions are suppressed on large scales and at late times. For a minimally-coupled massive scalar field $\sigma$, the two-point function has the exact solution\cite{Lee:2017okg}
\be
\label{massive-scalar-two-pt-fcn}
\langle  \sigma_{k} (\eta) \sigma_k (\eta')\rangle = \frac{\pi}{4} H^2 (\eta \eta')^{3/2} e^{- \pi \mu} H^{(1)}_{i \mu}(- k \eta) H^{(1)*}_{i\mu}  (- k\eta'),
\ee
where $\mu \equiv \sqrt{m/H^2 - 9/4}$, and $H^{(1)} _{i\mu}$ is the Hankel function of the first kind. This corresponds to a mode function $\sigma_k \sim H \eta^{3/2} e^{-\pi \mu/2} = a(\eta)^{-3/2}e^{-\pi \mu/2} /\sqrt{H}$, the latter equality using $\eta=-1/a H$ during inflation.

Now we turn to massive particles with spin. These are constrained by the Higuchi bound to have mass 
satisfying $m^2 \geq s (s-1) H^2$.  As for scalars, the isometries of dS fixes the scaling of the 
two-point correlation function of spinning fields \cite{AHM2015}, which takes the form
\be
\label{Os}
\langle {\cal O} _s (k) {\cal O}_{s} (k) \rangle \propto k^{2 \Delta - 3} ,
\ee
where all Lorentz indices are contracted with $s$ copies of a null vector, and $\Delta$ is the scaling 
dimension of the field,
\be
\Delta = \frac{3}{2} - i \mu_s \;\; ,\;\; \mu_s = \sqrt{\frac{m^2}{H^2} - 
\left(s-\frac{1}{2}\right)^2} .
\ee 
For heavy fields, or more precisely the ``principle series'' \cite{Lee:2017okg}, one has ${\rm Re} 
\Delta=3/2$, and the mode function is simply
\be
\langle {\cal O} _s (k) {\cal O}_{s} (k) \rangle \propto k^{ - 2 i \mu_s} .
\ee
The prefactors follow from dimensional analysis, and intuition from solving the 
Klein-Gordon equation for heavy fields, which leads to an additional $e^{-\pi 
\mu_s}$ suppression. Importantly, this applies to general operators with spin, and not to just to bosonic (integer spin) operators, but to fermionic (half-integer spin) as well, and the two-point function of half-integer operators  is similarly constrained to scale as $k^{2 \Delta -3}$ as in \eqref{Os}.

In this work we will focus on the angular dependence of correlation functions. Given this, for simplicity we ignore the $e^{i \mu_s}$ phase, though we note that this can 
lead to oscillations in $k$-space \cite{Lee:2016vti}, and thus is of potential interest. Dropping this phase, 
the mode functions for spin-$s$ and spin-$(s+1/2)$ are at late times given by 
 \be
{\cal O} _s (k,\eta)  \simeq \frac{a(\eta)^{-3/2}}{\sqrt{H}} e^{- \pi 
\mu_s/2}\;\; , \;\;  {\cal O} _{s+1/2} (k,\eta)  \simeq a(\eta)^{-3/2}e^{- \pi 
\mu_{s+1/2}/2}.
\ee
The scaling with  $a(\eta)$ follows from solving the mode-function equation of 
motion explicitly\footnote{In solving the Klein-Gordon equation explicitly, the scaling with $a(\eta)$ depends on the number of upper vs. lower indices, but this dependence cancels when all indices are contracted insider of 
correlation functions. For example, mode functions defined with respect to 
$\mathcal{O}_s$ with all lower indices are
\be
{\cal O} _{i_1.... i_s} (k,\eta)  \propto \frac{a(\eta)^{s-3/2}}{\sqrt{H}} 
\ee
See (B.76) and (C.7) of \cite{Lee:2017okg}: Using $\eta$=$1/(aH)$
and $N_{\lambda}$=$(1/\sqrt{k})(k/H)^{s-1}$, one finds that\\
$N_{\lambda}(k\eta)^{3/2 -s}$=$(1/\sqrt{k})(k/H)^{s-1}(k/aH)^{3/2-s}$
=$a^{s-3/2}/\sqrt{H}$. }, while the 
differing factors of $H$ follow from dimensional analysis\footnote{Bosons 
are defined as having mass term $m^2 ({\cal O}^b(x)) ^2$ and hence mass 
dimension $1$ while fermions are defined as having mass term $m \bar{\cal 
O}^f(x) {\cal O}^f(x)$ and hence mass dimension $3/2$. The corresponding mode 
functions have dimension $-1/2$ and $0$}. This matches with the known result for a heavy spin-$1/2$ particle in de Sitter space\cite{Lyth:1996yj}:  a 
spin-$1/2$ fermion $\chi$ with mass $m>H$ has mode function given by,
\be
\label{chik}
\chi_k \simeq  a(\eta)^{-3/2}e^{- \pi m_{\chi}/2H}.
\ee

Now we can make this more precise. A massive spin-s boson may be split into 
helicity components as,
\be
\sigma_{\mu_1...\mu_s} = \displaystyle \sum _{\lambda = -s} ^s \sigma^{\lambda} 
_{\mu_1...\mu_s},
\ee
and then decomposed into fields of $n$ polarization directions by projecting the 
spinning field $\sigma_{\mu_1...\mu_s}$ onto spatial slices, i.e. via the 
decomposition
\be
\sigma_{i_1...i_n\eta...\eta} = \sum_{\lambda} 
\sigma^\lambda_{n,s}\varepsilon^\lambda_{i_1...i_n},
\ee
where $\eta$ is the time coordinate. Here, the $s$ index refers to the spin, $n$ 
refers to the `spatial spin,' and $\lambda$ is the 
helicity of the field. Thus $\varepsilon^\lambda_{i_s...i_n}$ is a normalized, 
totally symmetric tensor with spin $s$ and helicity $\lambda$. The $\sigma_{n,s} ^{\lambda}$ satisfy $\sigma_{n,s} ^{\lambda}=0$ for $n<|\lambda|$ \cite{Lee:2016vti}.

The quantity that appears in scattering with scalars is $\lambda=0$ and $n=s$ (for more details see \cite{Lee:2016vti} or Appendix \ref{polvectors}), i.e. the quantity $\sigma^0 _{s,s}$.
Explicitly solving for the mode function, one finds \cite{Lee:2016vti}
\be
\sigma^0 _{s,s}(k,\eta)  \simeq \frac{a(\eta)^{s-3/2}}{\sqrt{H}} e^{- \pi 
\mu_s/2}.
\ee
Moreover, for the above $\lambda=0$ helicity state, one has the important 
relation,
\be
\label{Ps}
\hat{q}_{i_1}\hat{q}_{i_2}...\hat{q}_{i_s}\varepsilon^{\lambda=0}_{i_1...i_s 
}({\hat{k}}, \varepsilon) = P_s({
\rm} \cos\theta) ,
\ee 
with $\theta$ defined as the angle between $\hat{q}$ and $\hat{k}$. This follows from more 
general relations for spin-$s$ polarization vectors, which are detailed in the 
thesis \cite{Lee:2017okg}, and given in Appendix \ref{polvectors}.

Similar to the bosonic case, a massive spin-$(s+1/2)$ 4-component fermion may be 
split into helicity components as,
\be
\psi ^{\alpha} _{\mu_1...\mu_s} = \displaystyle \sum _{\lambda}  \psi^{\lambda 
\alpha} _{\mu_1...\mu_s},
\ee
where $\alpha$ is a fermionic index and $\mu_1 ... \mu_s$ are bosonic indices, 
and projected onto spatial slices via the decomposition,
\be
\psi ^{\alpha}_{i_1...i_n\eta...\eta} = \sum_{\lambda} 
\psi^\lambda_{n,s}\epsilon^{\lambda \, \alpha}_{i_1...i_n},
\ee
where again $\eta$ is the time coordinate.  We construct the spin-$(s+1/2)$ 
polarization vectors as a tensor product of spin-1/2 and spin-$s$. That is, we 
decompose,
\be
\epsilon^{\lambda \alpha} _{i_1...i_s} = \sum_{\lambda'} \xi_{\lambda'} 
^{\alpha} \varepsilon_{i_1...i_s} ^{\lambda} ,
\ee
with $\xi_{\lambda'}$ a spin-1/2 eigenspinor of helicity $\lambda'$ and 
$\varepsilon^{\lambda}$ the spin-$s$ polarization vector of helicity $\lambda$.  
The general decomposition of the fermion field can then be written as,
\be
\psi ^{\alpha}_{i_1...i_n\eta...\eta} = \sum_{\lambda, \lambda'} \psi^{\lambda 
\lambda'}_{n,s}\epsilon^{\lambda}_{i_1...i_n}  \xi^{\lambda' \alpha}.
\ee
Similar to the bosonic case, the fermions can be split into helicity states, and 
it is the helicity $0\pm1/2$ which contributes to the loop in Figure 
\ref{fig:fig1}. More explicitly, the relevant mode function has the form,
\be
\label{psi0ss}
\psi^{0 \lambda'} _{s,s} \simeq  a(\eta)^{s-3/2}e^{- \pi \mu_{s}/2}  .
\ee
One can use this,  along with the mode functions \eqref{chik} and 
\eqref{zetak}, and the interaction Lagrangian \eqref{LHS1f}, to compute 
correlation functions of $\zeta$ involving intermediate states of fermions.

\subsection{Non-Gaussianity from higher spin Particle Exchange}

The correlator we wish to compute is of three $\zeta(k,\eta)$ at lates times, 
$\eta \rightarrow 0$, and in the limit that one of the momenta is small $k_1 \ll 
k_{2},k_3$, i.e. the quantity,
\be
\langle \zeta(k_1,0) \zeta(k_2,0) \zeta(k_3,0) \rangle ,
\ee
given the insertion of interaction vertices between the scalar $\zeta$ and 
higher spin fields.  The result for exchange of a higher spin boson are given in 
\cite{AHM2015,Lee:2016vti}. While \cite{AHM2015} focused on the scaling and 
angular dependence, \cite{Lee:2016vti} explicitly solved the Klein-Gordon 
equation for higher spin bosonic fields and from this was able to compute all 
expressions exactly. In our analysis we will follow \cite{AHM2015} and focus on 
the amplitude and angular dependence.

The 3-point function resulting from higher spin boson exchange, diagram 
\ref{fig:fig2}, is given by\cite{Lee:2016vti},
\begin{align}
\lim_{k_1\ll k_3, \eta\rightarrow0} \frac{\langle\zeta(k_1)\zeta( k_2) \zeta(k_3)\rangle}{\Delta_\zeta^4}&\, = \,  
\alpha_s
 \Delta_\zeta^{-1}\times P_s(\hat{ k}_1\cdot\hat{k}_3)\times {\cal 
I}^{(s)}(\mu_s,c_\pi,k_1,k_3,k_3) \delta(\sum k_i) + ({k}_2\leftrightarrow { k}_3)\, ,
\label{eq:singleSq}
\end{align}
where $ {\cal I}^{(s)}(\mu_s,c_\pi,k_1,k_3,k_3)$ is a complicated function of momenta given in the Appendices of \cite{Lee:2016vti}, and $P_{s}$ the Legendre polynomial. This is characterized by a dimensionless coupling $\alpha_s$, which in the notation of our \eqref{HSbosonints}, and taking the Goldstone boson parameters of \cite{Lee:2016vti} to be $c_{\pi}=1$ and $f_{\pi}=m_{pl}$, is given by,
\be
\alpha_s = \lambda_s g_s\left(\frac{\Lambda}{m_{pl}}\right)^6 \left(\frac{H}{\Lambda}\right)^{2s+1}.
\ee
This corresponds to a non-Gaussianity parameter of
\be
f_{\rm NL}\equiv \frac{5}{18} \frac{\langle \zeta_k \zeta_k \zeta_k \rangle}{P_{\zeta}(k) ^2}\, \sim\, e^{-\pi \mu_s}  \alpha_s
 \Delta_\zeta^{-1}\, , \label{eq:fnlsingle}
\ee
with shape function
\be
\lim_{k_1 \ll k_3}\langle \zeta(k_1,0) \zeta(k_2,0) \zeta(k_3,0) \rangle \propto  \frac{1}{k_1^3 k_3^3}\left(\frac{k_1}{k_3}\right)^2\, 
, \label{analyticSq}
\ee
and a characteristic angular dependence,
\be
\langle \zeta(k_1,0) \zeta(k_2,0) \zeta(k_3,0) \rangle \propto P_{s}(\cos\theta),
\ee
 with $\theta$ the angle between $k_1$ and $k_3$. 

We now turn to the fermions. Before we proceed, it is important to further clarify and emphasize the procedure. We approximate the fermionic mode functions by their super-horizon scaling \eqref{psi0ss}, which neglects the sub-horizon oscillatory behavior of the exact solution (see \cite{Lee:2016vti} for the bosonic results). This is sufficient to compute the angular dependence of the 3-point function, as is our aim, but not the $k$-dependence and thus not the shape function. Additionally, we regulate the fermionic loop of Figure \ref{fig:fig1} by imposing a UV cutoff, which we choose to be the Hubble scale for self-consistency with our approximate form of the mode functions. The choice of cutoff does not qualitatively affect the result, and can be undone by a simple replacement $H\rightarrow \Lambda_{UV}$ in the loop integral.

To compute the fermion diagram Figure \ref{fig:fig1} there are two interaction Hamiltonians, which follow from the Lagrangian \eqref{LHS1f}, given by
\be
H_{int1} = \frac{\lambda_s}{\Lambda^{s-1}}  \int d^3x  \frac{1}{a^{2s-3}} \, 
\partial_{i_1 ... i_s} \zeta  \, \bar{\chi}  \, \psi_{ \, i_1 ... i_s}  + h.c.,
\ee
and
\be
H_{int2} =\frac{g_s}{\Lambda^{s}} \int d^3x  \frac{1}{a^{2s-2}} \,  
\zeta' \partial_{i_1 ... i_s} \zeta \, \bar{\chi} \, \psi_{ \, i_1 ... i_s} + h.c. ,
\ee
with $i$ indices summed over, and where $'$ indicates a derivative with respect to conformal time. The 3-point function of three $\zeta(k,\eta)$, at late times $\eta \rightarrow 0$, is given by 
\be
\langle \zeta(k_1,0) \zeta(k_2,0) \zeta(k_3,0) \rangle =  {\rm Re} 4 \langle 
\left[  \zeta(k_1,0) \zeta(k_2,0) \zeta(k_3,0) \displaystyle  \int a(\eta'){\rm d}\eta'  
H_{int 1}(\eta') \displaystyle \int a(\eta'') {\rm d}\eta''  H_{int 2}(\eta'')   \right] 
\rangle ,
\ee
where we focus on the second term in \eqref{eq:threept}, which has the same angular dependence as the first term, the two terms differing in calculation only by the distribution of complex conjugation among the resulting $\zeta$ mode functions. The additional factor of 2 as compared to \eqref{eq:threept} is due to the two possibilities of time-ordering $H_1$ and $H_2$. 

To compute this we expand $H_{int1}$ and $H_{int2}$ in momentum space, which results 
in 7 momentum integrals, ${\rm d}^3 q_1 .... {\rm d}^3 q_7$. We define the 
momenta as follows: let $k_1$ be the $\zeta$ in the left interaction vertex of 
Figure \ref{fig:fig1}, and $\ell$ the momentum of $\chi$ in the loop and  $\ell+ 
k_1$ the momentum of $\psi$ in the loop. The `outgoing' $\zeta$ momenta are 
defined as $k_2$ and $k_3$; under Wick contractions we will need to sum over 
$k_2 \rightarrow k_3$.

This gives, approximating the mode functions by their forms given in section \ref{sec:HSdS},
\bea
\label{zzz3}
&&\langle \zeta(k_1,0) \zeta(k_2,0) \zeta(k_3,0) \rangle =   \frac{\lambda_s 
g_s}{\Lambda^{2s-1}} {\rm Re}\,  8  \zeta_{k_1}(0)  \zeta_{k_2}(0)  \zeta_{k_3}(0)  
\int \frac{{\rm d} \eta'}{a(\eta')^{s-1}} \,  \int \frac{{\rm d} 
\eta''}{a(\eta'') ^{s}} \zeta_{k_1}^*(\eta') 
\zeta_{k_2}^{'*}(\eta'') \zeta_{k_3}^{*} (\eta'') \nonumber \\
&&\,\,\, \cdot |k_1|^s |k_3|^s e^{-  \pi \mu_s} e^{- \pi m_{\chi}/H}  \delta(\sum 
k_i)   \int \frac{{\rm d }^3 \ell}{(2 \pi)^3} \, P_{s}(\hat{k}_1 \cdot 
\hat{q}) P_{s}(\hat{k}_3\cdot\hat{q}) \left( \sum_{\lambda} 
\xi^{\lambda}(\hat{\ell}) \xi_\lambda ^\dagger (\hat{q})\right)^2 + k_2 \leftrightarrow 
k_3 ,
\eea
where $\vec{q}\equiv \vec{\ell} + \vec{k_1}$. We can factorize this into three 
pieces:
\be
\langle \zeta(k_1,0) \zeta(k_2,0) \zeta(k_3,0) \rangle \equiv {\rm Re} 
\mathcal{I}_1 \mathcal{I}_2 \mathcal{I}_3 ,
\ee 
with time-integrals,
\be
\mathcal{I}_{1} \equiv \zeta_{k_1}(0)  \zeta_{k_2}(0)  \zeta_{k_3}(0)  \int 
\frac{{\rm d} \eta'}{a(\eta')^{s-1}} \, \int \frac{{\rm d} \eta''}{a(\eta'') 
^{s}} \zeta_{k_1}^*(\eta') \zeta_{k_2}^{'*}(\eta'') \zeta_{k_3}^* (\eta'') ,
\ee
momentum integrals,
\be
\mathcal{I}_{2} \equiv \int \frac{{\rm d }^3 \ell}{(2 \pi)^3} \, P_{s}(\hat{k}_1 
\cdot \hat{q}) P_{s}(\hat{k}_3\cdot\hat{q}) \left( \sum_{\lambda} 
\xi^{\lambda}(\hat{\ell}) \xi_\lambda ^\dagger (\hat{q})\right)^2,
\ee
and an overall prefactor of 
\be
\mathcal{I}_3 \equiv \frac{\lambda_s g_s}{\Lambda^{2s-1}} 8  |k_1|^s |k_3|^s 
e^{-  \pi \mu_s} e^{- \pi m_{\chi}/H}  \delta(\sum k_i)  .
\ee

The loop integral $\mathcal{I}_2$ is UV-divergent, and we apply a cutoff at $H$,
\be
\mathcal{I}_2 \simeq H^3 \int {\rm d}\Omega \, P_s( \hat{k}_{1} \cdot \hat{q}) 
P_s( \hat{k}_{3} \cdot \hat{q}) \left( \sum_{\lambda} \xi^{\lambda}(\hat{\ell}) 
\xi_\lambda (\hat{q})\right)^2, 
\ee
with $\hat{q}$ now given by
\be
\hat{q} = \frac{H \hat{\ell} + k_1 \hat{k}_1}{\sqrt{H^2 + k_1 ^2}} \simeq 
\hat{\ell} ,
\ee
where the latter equality follows working in the limit $k_1 \rightarrow0$. This 
simplifies the sum over spin-1/2 helicities, as the $ \xi(\hat{k})$ are 
normalized to $1$. Thus we have,
\be
\mathcal{I}_2 \simeq H^3 \int \frac{{\rm d}\Omega}{(2 \pi^3)} \, P_s( 
\hat{k}_{1} \cdot \hat{q}) P_s( \hat{k}_{3} \cdot \hat{q}) .
\ee 
The remaining integral over angles can be performed analytically.  Defining 
$k_1$ as making angle $\theta_1=0$ in the $\{ x,y\}$ plane, and $k_3$ as making 
angle $\theta_{13}$, such that $\hat{k}_1 \cdot \hat{k}_3 = \cos \theta_{13}$, 
$\hat{q} \cdot \hat{k}_1 = \cos \theta$, $\hat{q} \cdot {k}_3 = \cos 
(\theta-\theta_{13})$, the integral can be written as,
\be
\mathcal{I}_2 \simeq \frac{H^3}{8 \pi^2} \int {\rm d}\cos \theta \, P_s( \cos 
\theta) P_s( \cos (\theta-\theta_{13})) .
\ee
We then use the identity\footnote{Using the spherical harmonics addition theorem 
with $\phi_1=\phi_2=0$.},
\be
P_{\ell} ({\rm cos}(a-b)) = \sum_{m=-\ell} ^\ell P_{\ell} ^m ({\rm cos} a)   
P_{\ell} ^m ({\rm cos} b) \frac{(\ell - m)!}{(\ell + m)!}  ,
\ee
from which one can evaluate the integral explicitly. The result is 
\be
\mathcal{I}_2 =  \frac{H^3}{8 \pi^2} \; \sum_{m=-s} ^s  c_m P_{s} ^m ( \cos 
\theta_{13})  ,
\ee
with coefficients
\be
\label{cm}
c_m =  \frac{(s - m)!}{(s + m)!}\displaystyle\int_{-1} ^1 {\rm d} x  \,  P_{s} 
(x) P_{s}^m (x) = \begin{cases}
    2 (-1)^{m/2} \frac{ s! (2s)!}{(1 + 2s)! (s+m)!},& \text{if } m\, {\rm 
even}\\
    0,              &{\rm if}\, m \, {\rm odd}  .
\end{cases}
\ee
Finally we can perform the time integration. Using the explicit $\zeta$ mode 
function \eqref{zetak}, one can analytically compute these integrals to find,
\be
\mathcal{I}_1 = \frac{ H^{2s+5}}{m_{pl}^6 (4 \epsilon)^3} \frac{1}{(k_1 k_2 
k_3)^3} (1+s) \Gamma(s) \Gamma(2+s) \frac{k_2 ^3}{k_1 ^{s} (k_2 + k_3)^{s+3}}  \left( 
1+ (s+3)\frac{k_3}{k_2}\right).
\ee
Putting the pieces together, we find for the non-Gaussianity, 
\bea
\label{zzz1}
&&\langle \zeta(k_1,0)  \zeta(k_2,0) \zeta(k_3,0) \rangle \simeq  
\frac{1}{64\pi} \lambda_s g_s  \frac{H^6}{m_{pl}^6 \epsilon^3}    (1+s) \Gamma(s) \Gamma(2+s)   \, 
\left( \frac{H}{\Lambda} 
\right)^{2s-1}  e^{-  \pi \mu_s} e^{- \pi m_{\chi}/H}   \nonumber \\
&& 
 \;\;\; \cdot \frac{\delta(\sum k_i)}{(k_1 k_2 k_3)^3}   \frac{k_3 ^s }{(k_2 + k_3)^{s}} \frac{k_2 ^3 H^3}{(k_2 + k_3)^3} \left( 
1+ (s+1)\frac{k_3}{k_2}\right)  \sum_{m=-s} 
^s  c_m P_{s} ^m ( \hat{k}_1 \cdot \hat{k}_3)   +  k_2 \leftrightarrow k_3  %~~ ,
\eea
which can be brought to a canonical form,
\be
\label{zzz2}
\langle \zeta(k_1,0) \zeta(k_2,0) \zeta(k_3,0) \rangle \simeq 
\mathcal{A}_{s+1/2}\frac{\Delta_{\zeta}(k) ^4}{k^6} \mathcal{S}(k_1,k_2,k_3)  
\delta(\sum k_i) \sum_{m=-s} ^s  c_m P_{s} ^m (\hat{k}_1 \cdot \hat{k}_3)  +  
k_2 \leftrightarrow k_3 ,
\ee
where $\Delta_{\zeta}^2$ is the 
dimensionless primordial power spectrum, ${\cal S}(k_1,k_2,k_3)$ is a  function 
of the ratios of $k_i$, and $\mathcal{A}$ is all remaining prefactors. 

The angular dependence of the non-Gaussianity is given by,
\be
\langle \zeta(k_1,0) \zeta(k_2,0) \zeta(k_3,0) \rangle \propto \sum_{m=-s} ^s  
c_m P_{s} ^m (\hat{k}_1 \cdot \hat{k}_3) .
\ee
with the coefficients $c_m$ given by \eqref{cm}. The schematic form of the shape function ${\cal S}$ can be read off from \eqref{zzz1}, but the exact 
expression requires solving for the exact mode-functions of the higher spin 
particles in de Sitter space.

The corresponding non-Gaussianity parameter is given by, 
\be
f_{NL} \simeq  \lambda_s g_s (1+s)  \Gamma(s) \Gamma(2+s)  \left( \frac{H}{\Lambda} \right)^{2s-1}  e^{- \pi 
\mu_s} e^{- \pi m_{\chi}/H} \Delta_{\zeta} ^2 ,
\ee
where the factor $\Gamma(s) \Gamma(2+s)$ is a relic of not having normalized the mode functions; we expect that as in the bosonic case, equation (A.77) of \cite{Lee:2016vti}, the normalization of the exact solution of the mode functions scales with $1/\Gamma(s)^{2}$, cancelling the $\Gamma(s)$ dependence of the 3-point function. The robust result is the scaling with the couplings $\lambda_s$, $g_s$, the ratio $H/\Lambda$, and the Boltzmann suppression due to both the higher-spin fermion and the spin-1/2 fermion, $e^{ - \pi \mu_s} e^{- \pi \mu_\chi}$. 

\subsection{The predictions of Higher Spin Supersymmetry}

We can now read-off result for the three-point function $\langle \zeta \zeta 
\zeta \rangle$ given the higher spin supermultiplet. We simply add 
the contributions from the particle content of the half-integer superspin 
$\Ysf=s+1/2$
supermultiplet \eqref{Yhalfint}, given the non-Gaussianity from each spin derived in the previous 
section. The result is
\bea
\langle \zeta(k_1,0) \zeta(k_2,0) \zeta(k_3,0) \rangle_{\rm HS-SUSY} =  
&&\langle \zeta(k_1,0) \zeta(k_2,0) \zeta(k_3,0) \rangle_{s+1} \nonumber \\
 && + 2 \times \langle \zeta(k_1,0) \zeta(k_2,0) \zeta(k_3,0) \rangle_{s+1/2} 
\nonumber \\
 && + \langle \zeta(k_1,0) \zeta(k_2,0) \zeta(k_3,0) \rangle_{s} \label{zzzHS} \\
  \propto && 
 P_{s+1}(\hat{k}_1 \cdot \hat{k}_3) \; ,\; \displaystyle \sum_{m=-s} ^s 
P_{s} ^m (\hat{k}_1 \cdot \hat{k}_3)\; ,\; P_{s}(\hat{k}_1 \cdot \hat{k}_3)
\eea
where the last line indicates that the three terms in \eqref{zzzHS} have angular dependence given by $P_{s+1}$, $\sum_m 
P_{s} ^m$, and $P_{s}$ respectively. The relative amplitudes are determined by the mass spectrum of the theory.

The quantitative amplitude of this signal is, as in the non-supersymmetric 
bosonic case \cite{Lee:2016vti}, generally small  $f_{NL} \lesssim \mathcal{O}(1)$. 
The primary obstruction making $f_{NL}$ any larger than this is perturbativity 
of the interaction strength, which at the very least, requires $\lambda_s 
(H/\Lambda)^{s-1} \ll 1$ and $g_s (H/\Lambda)^{s}\ll 1$, as these are the effective interaction strengths, e.g. appearing in \eqref{zzz3}. Non-Gaussianity of this size is not what would 
traditionally be referred to as `large', but it can be considerably larger than 
the slow-roll suppressed single-field slow-roll value, and is within reach for 
CMB-S4 \cite{Abazajian:2016yjj}.
 
The analysis can be repeated for the case of embedding the higher spin particles inside the integer superspin supermultiplet \eqref{Yint} instead of the half integer one we have used as an example. In that case the particle contained is $(~s+1/2,~s,~s,~s-1/2)$ therefore one can immediately read off the result. The known $P_{s}(\cos \theta)$ dependence of spin-$s$ bosons \cite{AHM2015,Lee:2016vti} is accompanied by two towers of associated Legendre polynomials, $\sum_m P_{s} ^m$ and $\sum_m P_{s-1} ^m$, from the $s+1/2$ and $s-1/2$ fermions respectively.

%%%%%%%%%%%%%%%%%%%%%%%%%%%%%%%%%%%%%%%%%%%%%%
\section{Discussion}
\label{outro}
%%%%%%%%%%%%%%%%%%%%%%%%%%%%%%%%%%%%%%%%%%%%%%

Precision measurements of the cosmic microwave background provide an 
unprecedented opportunity to search for new physics in the early universe. The 
3-point function of primordial curvature perturbations, $\langle \zeta \zeta 
\zeta \rangle$, colloquially referred to as the non-Gaussianity, is sensitive to 
any new degrees of freedom, including those that are naively too heavy to be 
excited. One of the most striking results of this research program is the 
non-Gaussianity due to higher spin particles, and in particular the angular 
dependence $\langle  \zeta \zeta  \zeta  \rangle \propto P_{s}(\cos \theta)$ due 
to the exchange of a single spin-$s$ boson \cite{AHM2015}. This prompted a 
flurry of activity, and possibilities for observing  \cite{Bartolo2018, 
Moradinezhad20181,MoradinezhadDizgah:2017szk,Moradinezhad20182,Bordin2019,Kogai:2018nse,Kehagias:2017cym,Franciolini:2017ktv,Franciolini:2018eno} 
the signature of higher spin particles.

Higher spin fermions have heretofore been left out of this discussion, but 
insofar as higher spin theory is understood as a limit of quantum gravity, 
namely superstring theory, fermions are built into the theory. This is required 
by the supersymmetric nature of the theory, which is itself a powerful tool for 
the incorporation of fermions into string theory, e.g. the construction of 
fermionic D-brane actions is accomplished by relying on the underlying 
supersymmetry of the theory 
\cite{Marolf:2003vf,Marolf:2003ye,Martucci:2005rb,Dasgupta:2016prs}.  Guided by 
this, and building on recent developments in the construction of supersymmetric 
higher spin theories 
\cite{hss2,hss3,hss4,BKS,GKS96a,GKS,hss5,hss6,hss7,hss8,mhs1,mhs2,mhs3,mhs4}, we 
have studied the imprint of higher spin supersymmetry at the cosmological 
collider.

The main result of this paper is a characteristic pattern of the angular 
dependence of $\langle \zeta \zeta \zeta \rangle$ due to the exchange of 
higher spin superpartners. We find the $P_{s}(\cos \theta)$ signature of 
higher spin boson exchange, with $\theta$ the angle between the short and long 
wavelength modes, comes along with a $P_{s+1}(\cos \theta)$ and a tower of 
associated Legendre polynomials, arising from a spin-$s+1$ boson and a pair of 
spin-$s+1/2$ fermions. For
a variant description of higher spin supermultiplet, the partner contributions
can be instead two towers of associated Legendre polynomials}.
The amplitude of the signal is generically not large by 
comparison to other  known sources (e.g. \cite{Chen:2013aj,Holman:2007na}), as 
already known for the non-supersymmetric bosonic case \cite{Lee:2016vti}, so it 
is indeed the angular dependence which gives this signal its elevated status. 
Given this, in this work we have not endeavored to do a rigorous and precise 
calculation of the shape-function, which requires explicitly solving the 
mode-functions and computing involved integrals \cite{Lee:2016vti}. This latter 
difficulty motivated the development of the cosmological \emph{bootstrap} 
\cite{Arkani-Hamed:2018kmz}, which might be a promising direction to take this 
work as well.

Remarkably, we have been able to derive these results despite not having a 
complete theory or model realization of higher spin supergravity inflation. 
Progress despite incomplete knowledge is a familiar situation in theoretical 
physics, for example, supersymmetry and the Green-Schwarz mechanism, all work to 
date pertaining to M-theory \cite{Witten:1995ex,Horava:1995qa}, or in a more 
recent context, Double Field Theory \cite{Hull:2009mi}. To overcome this, we 
have constructed an effective theory that combines higher spin supersymmetry 
with de Sitter supergravity and the effective field theory of inflation, to 
describe a higher spin sector minimally coupled to the inflationary sector such 
that the higher spin sector retains on-shell supersymmetry. This allowed us to 
use supersymmetry considerations to deduce the field content and interactions of 
the higher spin fields with the curvature perturbation.

There are a number of ways forward from here. We have not considered yet the 
interactions with the graviton, or for that the matter, the gravitino. The 
former of these, corresponding to primordial gravitational waves $\gamma$, 
itself can lead to an interesting three point function, $\langle \gamma \zeta 
\zeta \rangle$, probed by cross-correlation with CMB  B-mode polarization \cite{Meerburg:2016ecv}. 
Starting from an effective field theory guided guess for the relevant 
interaction,
\be 
\mathcal{L}_{\gamma HS} = \frac{\hat{\lambda}_s}{\Lambda^{s-2}}\partial_{{i_1}...i_{s-2}} 
\dot{\gamma}_{i_{s-1} {i_s}} \bar{\chi} \psi^{i_1...i_s}  + h.c. ,
\ee 
the $\langle \gamma \zeta \zeta \rangle$ computed can be straight-forwardly 
worked out in a fashion similar to section 4. We provide the calculations for 
this in Appendix \ref{gzz}, and here we give the result: 
\be 
\begin{split}
\lim_{k_1 \ll k_2,k_3}\langle \gamma^\lambda (k_1) \zeta (k_2) \zeta (k_3) \rangle = \hat{\lambda}_sg_s\left(\frac{H}{\Lambda}\right)^{2s-2}\frac{H^6}{m_{pl}^6}\frac{\sqrt{2}}{(4\epsilon)^2}  e^{-\pi 
m_\chi/H}e^{-\pi\mu_s} \delta(\sum k_i) \\ 
\cdot \mathcal{S}(k_1, k_2, k_3)  \int \frac{d\Omega}{(2 \pi)^3} P_{s-2} (\hat{k}_1 \cdot \hat{q}) \varepsilon^\lambda _{ij} (
\hat{k}_1) \, \sum_{\lambda'=\pm2}\varepsilon^{\lambda' ij} (\hat{q}) \hat{P}_{s}^{\lambda'} 
(\hat{k}_3 \cdot \hat{q})\mathcal{E}^{\lambda'}_2(\hat{k}_3\cdot\hat{q})  + k_2 \leftrightarrow k_3,
\end{split}
\ee 
where $\lambda=\pm2$ is the helicity of the external graviton, $\epsilon_{ij}$ is the spin-2 polarization tensor,  $\hat{P}_{s} ^\lambda(x) \equiv (1-x^2)^{-\lambda/2} P_{s} ^\lambda (x)$ and $\mathcal{E}_2^\lambda(\hat{k}_1\cdot \hat{k}_3) = 
\epsilon^\lambda_{ij}(\hat{k}_1)\hat{k}_3^i \hat{k}_3^j$ as in \cite{Lee:2016vti}, and we have put all $k$-dependence in the function $\mathcal{S}$. This result is characterized by an angular dependence that is an integral over 
 Legendre and associated Legendre polynomials,
\be 
\begin{split}
\langle \gamma^\lambda \zeta\zeta \rangle \propto  \sum_{\lambda'=\pm2} \int d\cos\theta_q\, P_{s-2} (\hat{k}_1 \cdot \hat{q}) \varepsilon^\lambda _{ij} (
\hat{k}_1) \varepsilon^{{\lambda'} ij} (\hat{q}) \hat{P}_{s}^{\lambda'} 
(\hat{k}_3 \cdot \hat{q})\mathcal{E}^{\lambda'}_2(\hat{k}_3\cdot\hat{q}) ,
\end{split}
\ee 
where $\theta_q$ is the angle $\hat{q}$ makes in the plane. In the supersymmetric 
context, this would be joined with contributions from additional
interactions. This requires careful consideration of the gravity multiplet, as is the focus of the supersymmetric EFT of inflation \cite{Delacretaz:2016nhw}. We postpone this analysis to future work.

On the theoretical front, an important next step is to construct the full theory 
of spontaneously broken supersymmetry (as in de Sitter supergravity and the supersymmetric EFT of inflation) and 
interacting higher spin fields. From this one can generalize the analysis here 
to situations where the higher spin fields themselves contribute to the 
supersymmetry breaking, or perhaps even drive inflation. We leave this 
possibility, and a host of observational implications, to future work.

As a concluding remark, we would like to express and share our enthusiasm
for the work of scientists who are searching for signals of supersymmetry
in the cosmos, a sentiment expressed by one of the authors 
in \cite{swSUSY}. As argued for in this work,
the non-Gaussianity of the CMB may
prove to be a powerful tool of discovery, and with some good fortune, 
perhaps more and different such tools will later emerge for the SUSY 
search at the Cosmic Collider.

\phantom{a}\hfill{``\emph{I want my @!\% !**\#\textasciitilde @\#  signal... where is it?  It's time to pay the piper!!}''}

\phantom{a}\hfill{-- B. Richter, in conversation overheard}\\
\phantom{a}\hfill{between the SLAC Director and}\\
\phantom{a}\hfill{physicist B.J. Bjorken.}

\vspace{1cm}

%%%%%%%%%%%%%%%%%%%%%%%%%%%%%%%%%%%%%%%%%%%%%%%
%%%%%%%%%% Acknowledgment %%%%%%%%%%%%%%%%%%%%

{\bf Acknowledgments}\\[.1in] \indent
This paper is dedicated to the memory of Burton Richter.
The authors thank Hayden Lee for many helpful and insightful discussions during the completion of this work.
The research of S.\ J.\ G.\ and K.\ K.\ is supported by the 
endowment of the Ford Foundation Professorship of Physics at 
Brown University. This work was partially supported by the U.S. National 
Science Foundation grant PHY-1315155.%\vspace{5ex}\\
\newpage

\appendix

\pagebreak
\setcounter{equation}{0}
\renewcommand{\theequation}{\thesection.\arabic{equation}}

\section[Spin s polarization vectors]{Spin-$s$ polarization vectors}
\label{polvectors}

This appendix discusses some relevant preliminaries and definitions relating to 
the free theory of higher spin fields in de Sitter space, which can also be 
found in \cite{Lee:2016vti}. Following \cite{Lee:2016vti}, vectors will be denoted here in boldface, e.g. ${\bf k}$.

  It is convenient to project the spinning field, 
$\sigma_{\mu_1...\mu_s}$ onto spatial slices, which we can then write as
\be
\sigma_{i_1...i_n\eta...\eta} = \sum_{\lambda} 
\sigma^\lambda_{n,s}\varepsilon^\lambda_{i_1...i_n}.
\ee
Here, the s index refers to the spin, n refers to the `spatial spin,' and 
$\lambda$ is the helicity of the field. $\varepsilon^\lambda_{i_s...i_n}$ is a 
normalized, totally symmetric tensor with spin s and helicity $\lambda$. It must 
satisfy: 
\be 
\label{polvectorprops}
\varepsilon^\lambda_{i_1...i_s} = \varepsilon^\lambda_{(i_1...i_s)} \;\;,\;\;
\varepsilon^\lambda_{iii_3...i_s} = 0\;\;\;, 
\;\;\;\hat{k}_{i_1}...\hat{k}_{i_r}\varepsilon^\lambda_{i_1...i_s} = 0 \;\;\mbox{for}\;\; r 
> s - |\lambda|,
\ee
corresponding to the symmetric, traceless, and transverse properties. These properties of the polarization tensor imply that we can decompose it into 
transverse and longitudinal parts as,
\be
\varepsilon^\lambda_{i_1...i_s}(\hat{\mathbf{k}}, \varepsilon) = 
\varepsilon^\lambda_{(i_1...i_\lambda}(\mathbf{\varepsilon})f_{i_{\lambda+1}...i_s)}(\hat{\mathbf{k}}),
\ee 
where $ \varepsilon^\lambda_{i_1...i_\lambda}$ is a maximally transverse polarization 
tensor, constructed out of polarization vectors $\varepsilon^{\pm}$ that are 
perpendicular to $\hat{\mathbf{k}}$. We must have that $\varepsilon^+ = 
(\varepsilon^-)*$, so that $ \varepsilon^\lambda_{i_1...i_\lambda}$ can be 
specified, up to a phase, by a single polarization vector ${\bf \varepsilon}$. 
We have also defined $f_{i_{\lambda+1}...i_s}$ as the longitudinal part of the associated Legendre 
polynomial, after contraction with momenta. 

We then define,
\be
F_{s} ^\lambda = q_{i_1}....q_{i_s} \varepsilon^{\lambda} _{{i_1}...{i_s}} (\bf k) ,
\ee
The symmetry properties of $\varepsilon$ imply that $F_{s} ^\lambda$ takes the form \cite{Lee:2016vti}, in d=3 spatial dimensions,
\be
\label{Fs}
F_{s} ^\lambda \propto z \hat{P} _s ^\lambda ,
\ee
where $z\equiv q_{i_1}...q_{i_\lambda} \varepsilon^\lambda_{i_{1}....i_\lambda}$, and $\hat{P}$ defined via
\be
P_{s}^{\lambda}(\theta,\phi)  = \sin ^{\lambda} \theta \hat{P}_s ^\lambda (\theta,\phi),
\ee
where $P_{s} ^\lambda$ is the associated Legendre polynomial. For the special case of $\lambda=0$, which appears in the calculation of 3-point functions after enforcing momentum conservation, the othonormality of differing helicity states $\lambda$ and $\lambda'$,  and the transverse property \eqref{polvectorprops}, one has
\be
q_{i_1}....q_{i_s} \varepsilon^{0} _{{i_1}...{i_s}}({\bf \hat{k}})  \propto  P_{s}( \hat{{\bf q}} \cdot \hat{{\bf k}}) ,
\ee
with magnitude $|q_1|^s$, leading to the characteristic angular dependence of the three-point function for spin-$s$ boson exchange. Moreover, the transverse property also implies that the only $\sigma^{0} _{n,s}$ that enters the correlation function is $n=s$.

\section[Details of gamma zeta zeta calculation]{Details of $\langle \gamma\zeta\zeta \rangle$ Calculation}
\label{gzz}

\begin{figure}[ht]
\begin{center}
\includegraphics[scale=0.4]{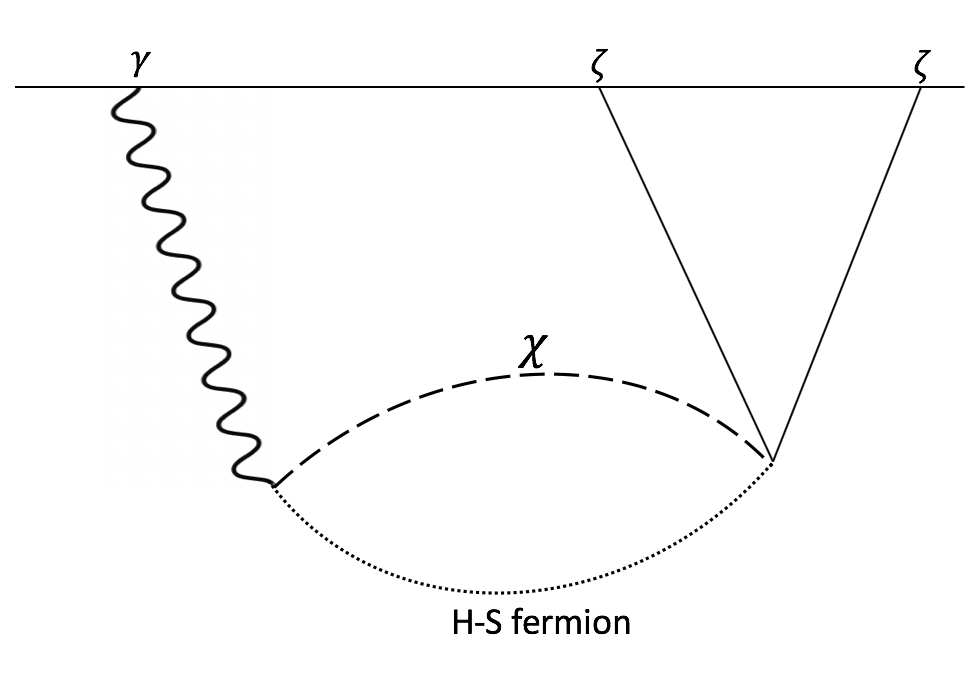}
\caption{Diagram contributing to $\langle \gamma\zeta\zeta \rangle$.}
  \label{fig:tssdiagram}
\end{center}
\end{figure}

\setcounter{equation}{0}
In this appendix, we further explicate the derivation of the tensor-scalar-scalar correlation function. We will limit our analysis to the single-exchange diagram shown in Figure \ref{fig:tssdiagram}.

This diagram has the same form as Figure \ref{fig:fig1}, however now we have an external graviton carrying momentum $k_1$ instead of $\zeta$. The relevant interaction Lagrangian we will consider is 
\be 
\mathcal{L} = 
\frac{\hat{\lambda}_s}{\Lambda^{s-2}}\partial_{{i_1}...i_{s-2}} 
\dot{\gamma}_{i_{s-1} {i_s}} \bar{\chi} \psi^{i_1...i_s} + 
\frac{g_s}{\Lambda^{s}}\dot{\zeta} \partial_{i_1...i_{s}}\zeta 
\bar{\chi}\psi^{i_1...i_s} + h.c. , 
\ee
which corresponds to two interaction Hamiltonians 
\be 
H_{int_1} = \frac{\hat{\lambda}_s}{\Lambda^{s-2}}\int d^3x 
\frac{1}{a^{2s-2}}\partial_{i_1...i_{s-2}}\gamma'_{i_{s-1}i_s}\bar{\chi} 
\psi^{i_1...i_s} +h.c., 
\ee
\be
H_{int_2} = \frac{g_s}{\Lambda^{s}}\int d^3x \frac{1}{a^{2s-2}}\zeta' 
\partial_{i_1...i_{s}}\zeta \bar{\chi} \psi^{i_1...i_s} + h.c.. 
\ee
As in the $\langle \zeta\zeta\zeta \rangle$ calculation, we would like to expand 
in Fourier modes. The graviton can be expanded in helicity modes as given in
\cite{Lee:2016vti}:
\be 
\gamma_{ij}(\mathbf{k}, \eta) = \sum_{\lambda=\pm2} 
\varepsilon^\lambda_{ij}(\mathbf{k})\gamma^\lambda_k(\eta)b(\mathbf{k}, \lambda) 
+ h.c., 
\ee
where the graviton mode function, $\gamma^\lambda_k$, is given by 
\be 
\gamma_k^\lambda(\eta) = \frac{\sqrt{2}H}{m_{pl}}\frac{1}{\sqrt{2k^3}}(1+ 
ik\eta)e^{-ik\eta}. 
\ee

With the mode functions in hand we can compute the tensor-scalar-scalar three point function $\langle\gamma \zeta\zeta\rangle$.  As before, we would like to expand each interaction Hamiltonian in momentum space and compute the correlator. Much of the calculation remains the same as in the $\langle\zeta\zeta\zeta\rangle$ case, however, there are some subtleties. The angular dependence due to $H_{int_2}$ remains largely the same, however now due to the fact that $\lambda \neq 0$, we have $H_{int_2} \propto  \mathcal{E}_2^\lambda(\hat{k}_3\cdot \hat{q}) \hat{P}^\lambda_s(\hat{k}_3 \cdot \hat{q})$, rather than simply $P_s(\hat{k}_3\cdot \hat{q})$ as in the  $\langle\zeta\zeta\zeta\rangle$ case. This arises from the definition \eqref{Fs}, and we have defined $\mathcal{E}_2^\lambda(\hat{k}_3\cdot \hat{q}) = 
\varepsilon^\lambda_{ij}(\hat{k}_3)\hat{q}^i \hat{q}^j$. The angular dependence of $H_{int_1}$ is similarly complicated due to the contraction with $\gamma_{i_{s-1},i_s}$.

After expanding in momentum space and following a similar procedure as in the 
$\langle\zeta \zeta\zeta\rangle$ calculation, we obtain 
\be
\begin{split}
\langle \gamma^\lambda \zeta \zeta \rangle = 8\frac{\hat{\lambda}_s 
g_s}{\Lambda^{2s-2}}{\rm 
Re} \gamma^\lambda _{k_1}(0)\zeta_{k_2}(0)\zeta_{k_3}(0)\int 
\frac{d\eta'}{a(\eta')^{2s-3}} \int\frac{d\eta''}{a(\eta'')^{2s-3}}\int \frac{d^3\ell}{(2\pi)^3}\\
\cdot \gamma^{\lambda *'}_{k_1}(\eta')\zeta^{*'}_{k_2}(\eta'')\zeta^*_{k_3}(\eta'')\bar{\chi}_\ell(\eta')\chi_\ell(\eta'')\sum_{\lambda'=\pm2} \psi^{\lambda'}_{s,s,\ell+k_1}(\eta')\bar{\psi}^{\lambda'}_{s,s,\ell+k_1}(\eta'') 
\\
\cdot |k_1|^{s-2} |k_3|^{s} P_{s-2}(\hat{k}_1 \cdot 
\hat{q})\varepsilon^\lambda_{ij}(\hat{k}_1)  \varepsilon^{\lambda' ij}(\hat{q}) \hat{P}_{s}^{\lambda'} (\hat{k}_3 \cdot \hat{q}) 
\mathcal{E}^{\lambda'}_2(\hat{k}_3\cdot\hat{q})   \delta(\sum k_i) , 
\end{split}
\ee 
where we have defined $\mathcal{E}$ as in \cite{Lee:2016vti} as 
$\mathcal{E}_2^\lambda(\hat{k}_1 \cdot \hat{k}_3) = 
\epsilon^\lambda_{ij}(\hat{k}_1)\hat{k}_3 ^i \hat{k}_3 ^j$, and where $\vec{q}\equiv \vec{\ell} + \vec{k_1}$. The mode functions for 
$\chi$ remain the same as in the previous calculation, given by \eqref{chik},  
and we approximate that $\psi^{\pm2}_{s,s}$ by their super-horizon scaling, which is the same as \eqref{psi0ss}. Plugging in the explicit expressions for the mode functions 
and substituting $a(\eta) = -\frac{1}{H\eta}$, we have ,
\be 
\begin{split}
\langle \gamma^\lambda \zeta \zeta \rangle =\frac{\hat{\lambda}_sg_s}{\Lambda^{2s-2}} 
H^{2s}\frac{H^6}{m_{pl}^6} \frac{2\sqrt{2}}{(4\epsilon)^2} 
\frac{1}{k_1^3k_2^3k_3^3}e^{-\pi 
m_\chi/H}e^{-\pi\mu_s}|k_1|^{s-2}|k_3|^{s}\delta(\sum 
k_i)\\
\cdot{\rm Re}\int d\eta' \eta'^{s} (k_1^2\eta')e^{ik_1\eta'}\int d\eta'' 
\eta''^{s}(k_2^2\eta'')(1+ik_3\eta'')e^{-i(k_2 + k_3)\eta''} \\
\int \frac{d^3\ell}{(2\pi)^3} P_{s-2}(\hat{k}_1 \cdot 
\hat{q})\varepsilon^\lambda_{ij}(\hat{k}_1) \sum_{\lambda'=\pm2}  \varepsilon^{\lambda' ij}(\hat{q}) \hat{P}_{s}^{\lambda'} (\hat{k}_3 \cdot \hat{q}) 
\mathcal{E}^{\lambda'}_2(\hat{k}_3\cdot\hat{q}) + k_2 \leftrightarrow k_3 . 
\end{split} 
\ee 
where $\hat{P}_{s} ^\lambda(x) \equiv (1-x^2)^{-\lambda/2} P_{s} ^\lambda (x)$ as in \cite{Lee:2016vti}.

For ease of notation, let us denote this as,
\be 
\langle \gamma^\lambda \zeta \zeta \rangle = \mathcal{J}_1 \mathcal{J}_2\mathcal{J}_3 , 
\ee 
where $\mathcal{J}_1$ is the prefactor, $\mathcal{J}_2$ are the time integrals 
and $\mathcal{J}_3$ is the momentum integral. Performing the time integration 
and keeping only the real part yields 
\be
\mathcal{J}_2 = \frac{k_1^{-s} k_2^2 (k_2 + k_3(s+3))\Gamma(s+2)^2}{(k_2+k_3)^{s+3}}.
\ee 
In $\mathcal{J}_3$ we can perform the integration over $\ell$, enforcing a cutoff at $H$, leaving only the 
angular integral. Putting everything together, we obtain,
\be
\begin{split}
\lim_{k_1 \ll k_2,k_3}\langle \gamma^\lambda (k_1) \zeta (k_2) \zeta (k_3) \rangle = 
\frac{\hat{\lambda}_sg_s}{\Lambda^{2s-2}}\frac{H^{2s+6}}{m_{pl}^6} 
\frac{2\sqrt{2}}{(4\epsilon)^2} e^{-\pi 
m_\chi/H}e^{-\pi\mu_s} \delta(\sum k_i) 
\\
\cdot \frac{H^3 k_1^{-2}k_2^3k_3^{s}}{(k_1k_2k_3)^3(k_2+k_3)^{s+3}} \left(1 + \frac{k_3}{k_2}(s+3)\right)\Gamma(s+2)^2\\
\cdot \int \frac{d\Omega}{(2 \pi)^3} P_{s-2} (\hat{k}_1 \cdot \hat{q}) \varepsilon^\lambda _{ij} (
\hat{k}_1) \varepsilon^{\lambda ij} (\hat{q}) \hat{P}_{s}^\lambda 
(\hat{k}_3 \cdot \hat{q})\mathcal{E}^\lambda_2(\hat{k}_3\cdot\hat{q}) + k_2 \leftrightarrow k_3. 
\end{split}
\ee 
In a canonical form we can write this as,
\be 
\begin{split}
\lim_{k_1 \ll k_2,k_3}\langle \gamma^\lambda (k_1) \zeta (k_2) \zeta (k_3) \rangle = \hat{\lambda}_sg_s\left(\frac{H}{\Lambda}\right)^{2s-2}\frac{H^6}{m_{pl}^6}\frac{2\sqrt{2}}{(4\epsilon)^2}  e^{-\pi 
m_\chi/H}e^{-\pi\mu_s} \delta(\sum k_i) \\ 
\cdot \mathcal{S}(k_1, k_2, k_3) \sum_{\lambda'=\pm2} \int \frac{d\Omega}{(2 \pi)^3} P_{s-2} (\hat{k}_1 \cdot \hat{q}) \varepsilon^\lambda _{ij} (
\hat{k}_1) \varepsilon^{\lambda' ij} (\hat{q}) \hat{P}_{s}^{\lambda'} 
(\hat{k}_3 \cdot \hat{q})\mathcal{E}^{\lambda'}_2(\hat{k}_3\cdot\hat{q}) + k_2 \leftrightarrow k_3.
\end{split}
\ee 
The angular dependence is given by 
\be 
\begin{split}
\langle \gamma^\lambda \zeta\zeta \rangle \propto  \sum_{\lambda'=\pm2} \int d\cos\theta_q\, P_{s-2} (\hat{k}_1 \cdot \hat{q}) \varepsilon^\lambda _{ij} (
\hat{k}_1) \varepsilon^{{\lambda'} ij} (\hat{q}) \hat{P}_{s}^{\lambda'} 
(\hat{k}_3 \cdot \hat{q})\mathcal{E}^{\lambda'}_2(\hat{k}_3\cdot\hat{q}) ,
\end{split}
\ee 
where $\theta_q$ is the angle $\hat{q}$ makes in the plane.

%%%%%%%%%%%%%%%%%%%%%%%%%%%%%%%%%%%%
%%%%%%%%% Bibliography %%%%%%%%%%%%%%%%%%%%
%\printbibliography
%\small{
\bibliographystyle{unsrturl}
\bibliography{references}
%}

%\begin{thebibliography}{66}
%\small
%
%
%
%\end{thebibliography}

\end{document}